\documentclass[12pt,a4paper,fleqn]{article}
\usepackage{booktabs,latexsym}
\usepackage{graphicx,cite,amsmath,subfigure,epsfig,lscape,longtable,amssymb}
\usepackage{times,mathptm}
\usepackage{xcolor}

\newcommand {\di} {\displaystyle}
\newcommand{\p}{\partial}

\makeglossary

\title{\vspace*{-2cm}
Computational study for investigating acoustic streaming and heating during acoustic hemostasis.}

\author{\bf\normalsize
Maxim A. Solovchuk $^{1}$\thanks{\scriptsize solovchuk@gmail.com }
\hspace*{5pt}
Marc Thiriet $^{2}$
\hspace*{5pt}
Tony W. H. Sheu $^{1,3}$\thanks{\scriptsize twhsheu@ntu.edu.tw.
  }
\\
\footnotesize\sl $^1$ Center of Advanced Study in Theoretical Sciences (CASTS), National Taiwan University\\ [-1mm]
\footnotesize\sl $^2$ Sorbonne Universities, UPMC Univ Paris 06, UMR 7598, Laboratoire Jacques-Louis
Lions, F-75005, Paris, France \\[-1mm]
\footnotesize\sl $^3$ Department of Engineering Science and Ocean Engineering, National Taiwan University,\\[-4mm]
\footnotesize\sl      No. 1, Sec. 4, Roosevelt Road, Taipei, Taiwan 10617, Republic of China \\[-1mm]
\date{\empty}{\footnotesize\bf\today}}

\begin{document}
\baselineskip=22pt
\normalsize
\noindent
\rm

\maketitle

\abstract{High intensity focused ultrasound (HIFU) has many applications ranging from thermal ablation of cancer to hemostasis. Although focused ultrasound can seal a bleeding site, physical mechanisms of acoustic hemostasis are not fully understood yet. To understand better the interaction between different physical mechanisms involved in hemostasis a mathematical model of acoustic hemostasis is developed. This model comprises the nonlinear Westervelt equation and the bioheat equations in tissue and blood vessel. In the three dimensional domain, the nonlinear hemodynamic equations are coupled with the acoustic and thermal equations. Convected cooling and acoustic streaming effects are incorporated in the modeling study. Several sonication angles and two wound shapes have been studied. The optimal focal point location is at the rear of the wound and the optimal angle is $45^0$.
}

\begin{description}
\item[{\bf Keywords:}]{\it acoustic hemostasis; HIFU; Navier-Stokes equations; acoustic streaming; Westervelt equation}
\end{description}

\section{Introduction} \label{sec:introduction}
\noindent

Bleeding is one of the major causes of death after traumatic
injuries  \cite{Zderic2006}. Management of this type of injuries in 1987 in the USA, for example, accounted for \$64.7 billion (in 1993 dollars) \cite{VaezyZderic2007IJH,MillerLestina1987}.
Hemorrhage is stopped by vessel ligation, clamping,
and repair of the vessel \cite{Austin1995,Perry1995}.

Acoustic hemostasis is a new field of ultrasound research.
Focused
ultrasound has been successfully applied to the treatment of
tumors in different areas of the bodies, including the breast,
prostate, uterine fibroids and liver \cite{Zhou2011,Leslie2007}.
Other promising applications of high intensity focused ultrasound (HIFU) include small blood vessel
occlusion
\cite{HynynenColucci1996,SerroneBurgess2012,HendersonLewis2010},
hemostasis of bleeding vessels and organs
\cite{VaezyMartinJVS1999,VaezyMartin2001,Vaezy1997}
by delivering a large amount of acoustic energy to the bleeding site.

Although numerous experimental studies have been performed, the
optimal strategy for the acoustic hemostasis is not very clear.
The ultrasound frequency, intensity, duration of treatment,
location of focal point differ a lot in different experimental
studies. The focus can be fixed at the center of the wound, its proximal or its distal part.  Ultrasound focus can be also moved continuously along the
wound. Multiple sonications at several points can be used as well. Several studies have
been performed for punctured blood vessel
\cite{Zderic2006,VaezyMartin2001,VaezyMartinUMB1998}. Under the
same ultrasound parameters, the treatment time can differ even by an
order of magnitude for similar punctures
\cite{VaezyMartinUMB1998}. The difference can be attributed to the wound shape and the guidance of HIFU. Ultrasound
beam should be precisely located on the wound. If the location of
the focal point was not strictly controlled, a very large increase in
treatment time was required.
 In
order to improve the treatment planning processes involved during acoustic hemostasis should be understood.

Most of the acoustic hemostasis studies have been performed experimentally. There were only a few numerical studies \cite{TejaPanAmerican2013,SekinsAIP2012,Huang04}. In Ref. \cite{TejaPanAmerican2013} ultrasound propagation through different coupling materials between the tissue and transducer was studied. Heating in homogeneous media was considered. \cite{Huang04} studied the effects of blood flow cooling and acoustic streaming on ultrasound heating. Low intensity ultrasound with peak pressures up to 2 MPa was considered. They assumed a classical thermoviscous medium in which the absorption increases as frequency squared. For tissues the power law for absorption is more close to the linear dependence on frequency $f^1$. For high ultrasound powers and large peak pressures used in acoustic hemostasis applications this assumption is not valid. In order to take into account the correct absorption law, relaxation effect should be taken into account \cite{SolovchukSheuJASA}.
Recently, a three dimensional model for the determination of the influences of blood flow and the acoustic streaming on the temperature distribution was presented \cite{SolovchukSheuJASA,SolovchukSheuATE2013}. The proposed model was applied to get the temperature elevation in liver tumor in a patient specific geometry \cite{SolovchukJCS2014}. In the present paper the developed mathematical model will be applied to the acoustic hemostasis. Importance on the thermal and acoustic streaming
effects will be addressed.

Numerous experimental studies \cite{VaezyMartinUMB1998,VaezyMartin2001,VaezyNoble2004} on acoustic hemostasis have provided strong evidence that the thermal effect of focused ultrasound is responsible for the hemostasis. The absorbed ultrasound energy in tissues is transformed into the thermal energy during focused therapy and this energy deposition can quickly cause the tissue temperature to increase. Temperature elevation in excess of 70$^0$C in about 1 second \cite{VaezyMartin2001} allows sealing of the bleeding site. Blood flow at the wound can in theory carry away part of the deposited energy making the temperature elevation at the wound more complicated. In the
experiments \cite{VaezyMartinUMB1998} when the puncture site was exposed to the air, a jet of blood appeared out of the artery
after the puncture.  Such a jet flow can be stopped after focused ultrasound energy being applied to the bleeding site. In most of the experimental studies ultrasound beam was oriented perpendicularly to the wound. In the present paper several sonication angles and two wound shapes have been studied. We will show that sonication angle  and focal point location can be optimized in order to decrease the amount of bleeding.

\section{Methods} \label{sec:Three-field coupling}
\noindent

The three-dimensional (3D) acoustic-thermal-
hydrodynamic coupling model has been proposed to compute the
pressure, temperature, and blood flow velocity \cite{SolovchukSheuJASA}. The mathematical model \cite{SolovchukSheuJASA, SolovchukSheuATE2013} relies on a coupling of: (1)
nonlinear Westervelt equation with relaxation effects being
taken into account; (2) heat equations in biological tissues;
and (3) acoustic streaming hydrodynamic equations.

\subsection{ Nonlinear acoustic equation} \label{sec:acoustic}
Acoustic field generated by a HIFU source was modeled using
the coupled system of two partial differential equations given below \cite{SolovchukSheuJASA}
\begin{equation}
\label{WestSystem}
\begin{array}{l}
 \di  \nabla ^2p-\frac{1}{c_0 ^2}\frac{\partial
^2p}{\partial t^2} + \frac{\delta }{c_0 ^4}  \frac{\partial
^3p}{\partial t^3} + \frac{\beta }{\rho _0 c_0 ^4}\frac{\partial
^2p^2}{\partial t^2} +  \sum_i P_i =0
  \ \\
 \di (1+\tau_i \frac{\p}{\p t}) P_i = \frac {2}{c_0 ^3}  c_i \tau_i
  \frac{\partial
^3p}{\partial t^3} \ \\
 \end{array}
\end{equation}
The above system of equations
takes into account effects of diffraction, absorption, nonlinear propagation and relaxation effects. In the above,  $p$ is the sound pressure, $\beta
=1+\frac{B}{2A}\quad $ the coefficient of nonlinearity, and
$\delta $ the diffusivity of sound originating from fluid viscosity
and heat conduction, $\tau_i$ the relaxation time and
$c_i$ the small signal sound speed increment for the $i$-th
relaxation process. The first two terms describe the contribution of linear
lossless wave propagating at a small-signal sound speed. The third
term denotes the loss resulting from thermal conduction and fluid
viscosity. The fourth term accounts for acoustic nonlinearity which
may considerably affect thermal and mechanical changes within the
tissue. The last term models the inevitable relaxation
processes. In the present paper two relaxation processes ($i=2$) were
considered. All the unknown relaxation parameters shown above were calculated through
the minimization of a mean square error between the linear attenuation law
and the relaxation model \cite{SolovchukSheuJASA}.

For the linear Westervelt equation the intensity is equal to $I_L=p^2/2 \rho c_0 $.
For the nonlinear case the total intensity is
\begin{equation}
\label{I_Nonlin}
I=\sum_{n=1}^{\infty} I_n,
\end{equation}
where $I_n$ are the corresponding intensities for the respective
harmonics $n f_0$. The ultrasound power deposition per unit volume
is calculated by
\begin{equation}
\label{Q}
q=\sum_{n=1}^{\infty} 2 \alpha (nf_0) I_n
\end{equation}
The absorption coefficient in tissue shown above obeys the following frequency law:
\begin{equation}
\label{Alpha_tissue}
\alpha = \alpha_0 \left(\frac {f}{f_0} \right)^{\eta} ,
\end{equation}
where  $\alpha_0=8.1$ Np/m, $\eta = 1.0$ and $f_0=1$ MHz \cite{Duck1990}.



\subsection{ Energy equation for tissue heating}
 In the current simulation study of thermal field the
physical domain has been split into the domains for the perfused
tissue and the flowing blood. In a region free of large blood vessels,
the diffusion-type Pennes bioheat equation  \cite{Pennes1948} given below is employed to
model the transfer of heat in the perfused tissue region
\begin{equation}
   \label{BHTE1}
   \rho_t\:c_t\frac{\partial T}{\partial t}=k_t\nabla^2T-w_b\:c_b\left(T-T_\infty\right)+q
\end{equation}
In the above bioheat equation proposed for the modeling of
time-varying temperature in the tissue domain, $\rho$, $c$, $k$
denote the density, specific heat, and thermal conductivity,
respectively. The subscripts $t$ and $b$ refer to the tissue and
blood domains. The notation $T_\infty$ is denoted as the
temperature at a remote location. The variable $w_b~(\equiv 0.5$
kg/m$^3$-s) in Eq. (\ref{BHTE1}) is the perfusion rate for the
tissue cooling in capillary flows.

In the region containing large vessels, within which the blood
flow can convect heat, the biologically relevant heat source,
which is $q$, and the heat sink, which is
$-\rho_b\:c_b\:\underline{\mathbf{u}}\cdot\nabla T$, are added to
the conventional diffusion-type heat equation
\begin{equation}
   \label{BHTE2}
   \rho_b\:c_b\frac{\partial T}{\partial t}=k_b\nabla^2T-\rho_b\:c_b\:\underline{\mathbf{u}}\cdot\nabla T+q
\end{equation}
In the above, $\underline{\mathbf{u}}$ is the blood flow velocity.
We note here that the above
thermal equations (\ref{BHTE1},\ref{BHTE2}) are coupled with the acoustic equations
(\ref{WestSystem}) for the acoustic pressure through the power
deposition term $q$ defined in Eq. (\ref{Q}).

\subsection{ Acoustic streaming hydrodynamic equations} \label{sec:acoustic2}
Owing to the inclusion of heat sink shown on the right
hand side of Eq. (\ref{BHTE2}), the blood flow velocity plus
the velocity generated from the acoustic streaming due to the
applied high-intensity ultrasound must be determined. In this
study the flow in large blood vessels is assumed to be
incompressible and laminar. The vector equation for modeling the
blood flow motion, subject to the divergence free equation
$\nabla\cdot\underline{\mathbf{u}}=0$, in the presence of acoustic
stress vector is as follows \cite{SolovchukSheu2012,Kamakura1995}
\begin{equation}
   \label{Navier-Stoke}
   \frac{\partial\underline{\mathbf{u}}}{\partial t}
  +(\underline{\mathbf{u}}\cdot\nabla)\underline{\mathbf{u}}
  =\frac{\mu}{\rho}\nabla^2\underline{\mathbf{u}}-\frac{1}{\rho}\nabla{\mathbf{P}}+\frac{1}{\rho}\underline{\mathbf{F}}
\end{equation}
In the above, $\mathbf{P}$ is the static pressure,
$\mu$ ($=0.0035$ kg/m s) the shear viscosity of blood flow,
and $\rho$ the blood density.
In Eq. (\ref{Navier-Stoke}),
the force vector $\underline{\mathbf{F}}$ acting on blood fluid due to an incident ultrasound is assumed to
act along the acoustic axis $\underline{\mathbf{n}}$ and has the following form \cite[Ch. 7]{Hamilton1998}
\begin{equation}
 \label{F}
   \underline{\mathbf{F}}\cdot\underline{\mathbf{n}}=  -\frac {1}{c_0} \nabla \vec{I}=\frac {q}{c_0}
\end{equation}

\subsection{Solution procedure and description of the problem}

Nonlinear Westervelt equation (\ref{WestSystem}) is solved by finite difference method presented in Ref. \cite{SolovchukSheuJASA}. Discretization of this system of differential equations is started with the approximation of the temporal derivative $\frac {\p }{\p t} P_i^{n+1}$ shown in the second equation of the system (\ref{WestSystem}):
\begin{equation}
\label{West2Num}
 \di\ \frac{\p}{\p t} P_i^{n+1} = \frac {1}{2\Delta t} (3 P_i^{n+1}-4P_i^n +P_i^{n-1})\\
\end{equation}
After some algebraic manipulation the second equation in the system (\ref{WestSystem}) can be rewritten in the form:
\begin{equation}
\label{Pnu}
P_i^{n+1} = \frac {2}{c_0 ^3}  \frac { c_i \tau_i}{1+1.5 \tau_i/\Delta t } \frac {\partial
^3 p^{n+1}}{\partial t^3} - \frac{\tau_i}{2 \Delta t + 3 \tau_i } (-4P_i^n +P_i^{n-1})
\end{equation}
$P_{\nu}^{n+1} $ is then substituted into the first equation of the system (\ref{WestSystem}). The resulting equation
 will be solved implicitly.

Temporal derivatives in Westervelt equation are approximated using the following second order accurate schemes:
\begin{equation}
\label{ptt} \left. {\frac{\partial ^2p}{\partial t^2}}
\right|^{n+1}=\frac{2p^{n+1}-5p^n+4p^{n-1}-p^{n-2}}{(\Delta t)^2}
\end{equation}
\begin{equation}
\label{pttt} \left. {\frac{\partial ^3p}{\partial t^3}}
\right|^{n+1}=\frac{6p^{n+1}-23p^n+34p^{n-1}-24p^{n-2}+8p^{n-3}-p^{n-4}}{2(\Delta
t)^3}
\end{equation}
The nonlinear term $ \di \frac{\p^2 p^2}{\p t^2}\mid^{n+1}\ $ is linearized using the second order accurate relation:
\begin{equation}
\label{nonlin}
\begin{array}{l}
 \di \left. {\frac{\partial ^2p^2}{\partial t^2}} \right|^{n+1}=\frac{\partial
}{\partial t}\left. \left( {\frac{\partial p^2}{\partial t}}
\right) \right|^{n+1}=2\frac{\partial }{\partial t}\left( {p^n\left.
{\frac{\partial p}{\partial t}} \right|^{n+1}+p^{n+1}\left.
{\frac{\partial p}{\partial t}}
\right|^n-p^n\left. {\frac{\partial p}{\partial t}} \right|^n} \right)=  \ \\
 =2\left( {2p_t ^np_t ^{n+1}+p^np_{tt} ^{n+1}+p^{n+1}p_{tt} ^n-(p_t
^n)^2-p^np_{tt} ^n} \right)  \\
 \end{array}
\end{equation}
The above equations are then substituted into the Westervelt equation
to get the Helmholtz equation. This equation is then solved using the three-point sixth-order accurate scheme \cite{SolovchukSheuJASA}. Accuracy of the numerical solutions was examined in \cite{SolovchukSheuJASA} by comparing them with the known analytical and numerical
solutions obtained by other authors \cite{ONeil1949,Blackstock1966}.  Good agreement between the measured and
numerical results was also obtained \cite{SolovchukSheuJASA,SolovchukSheuATE2013}.

First the acoustic pressure was calculated. The acoustic pressure
was calculated only once for a given set of transducer parameters.
Afterwards, ultrasound power deposition in Eq. (\ref{Q}) and
acoustic streaming force in Eq. (\ref{F}) were determined and
stored. Blood flow velocity was computed from Eq. (\ref{Navier-Stoke}) at
every time step with the acoustic streaming effect being taken into
account and then was substituted to the bioheat equation (\ref{BHTE2}). With the
known blood flow velocities and power deposition terms, temperatures in blood flow domain and in tissue were calculated.
Initially, temperature is considered to be equal to $37^0$C. Temperature
continuity at the fluid-solid interface is imposed as that being applied in
a conjugate heat transfer problem. The interface boundary condition takes into account the thermal conduction in tissue and convection
in blood vessel domain. The three-dimensional problem is analyzed using finite-volume method. A detailed description of the solution procedures can be found in our previous articles \cite{SolovchukSheuJASA,SolovchukSheu2012,SolovchukSheuATE2013}.

The present 3D computational model was validated by comparing our simulated results for the temperature
field, with and without flow, with the experimental results of Huang et al. \cite{Huang04}. The  computational model for the prediction of acoustic streaming field was validated by comparing the results with those of Kamakura et al. \cite{Kamakura1995}. Temperature elevation by HIFU in ex-vivo porcine muscle was studied experimentally as well by MRI and numerically \cite{SolovchukMRI2014}. We demonstrated that for peak temperatures below 85-90$^{\circ} C$ our numerical simulation results are in excellent agreement with the experimental data measured in three dimensions. Both temperature rise and lesion size can be well predicted. For peak temperatures above 85-90$^{\circ} C$ preboiling or cavitation activity appears and, by consequence, lesion distortion starts, causing a small discrepancy between the measured and simulated temperature rises.

\begin{figure}
\centering
\mbox{\subfigure[]{\includegraphics[width=0.35\textwidth]{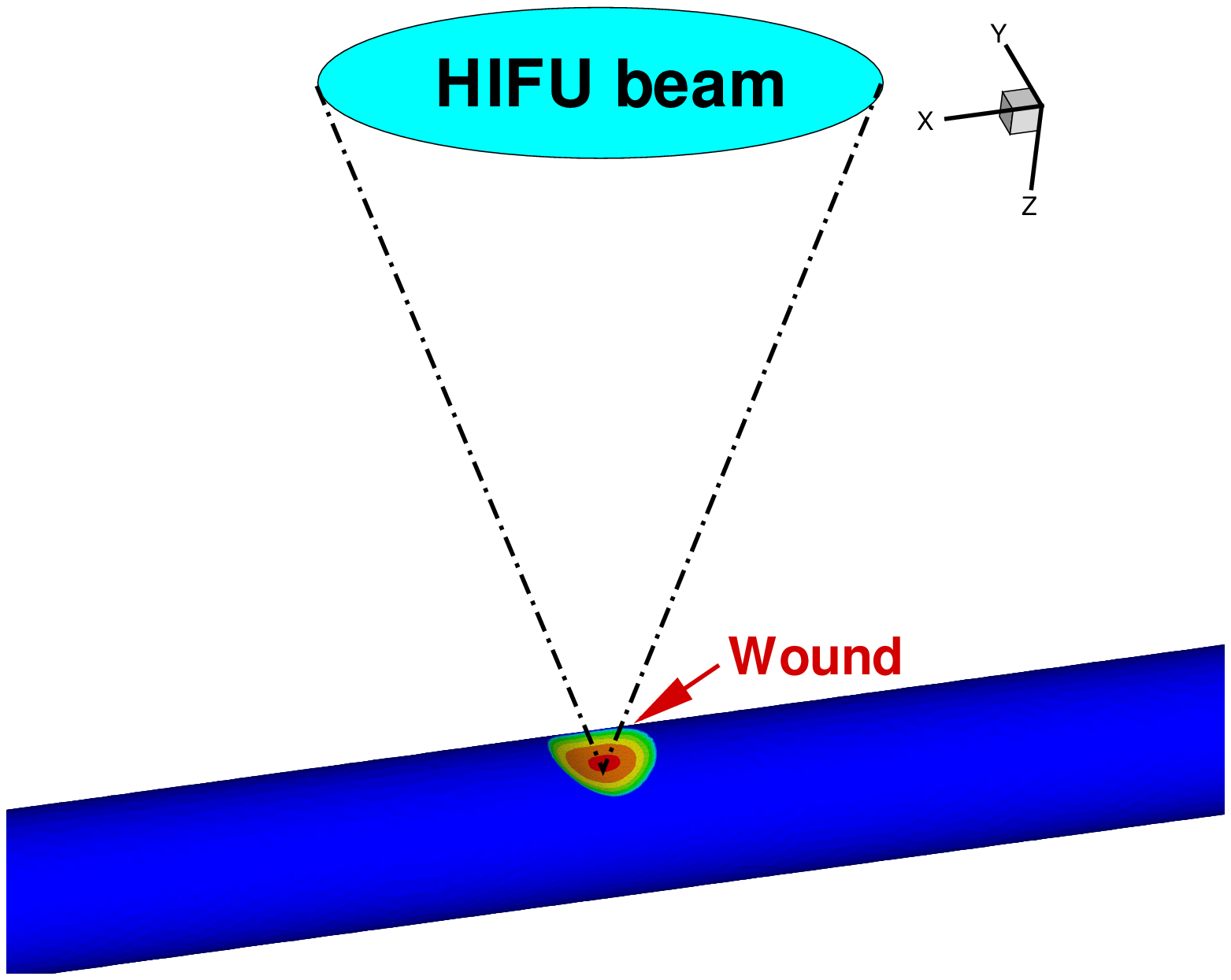}}
      \subfigure[]{\includegraphics[width=0.4\textwidth]{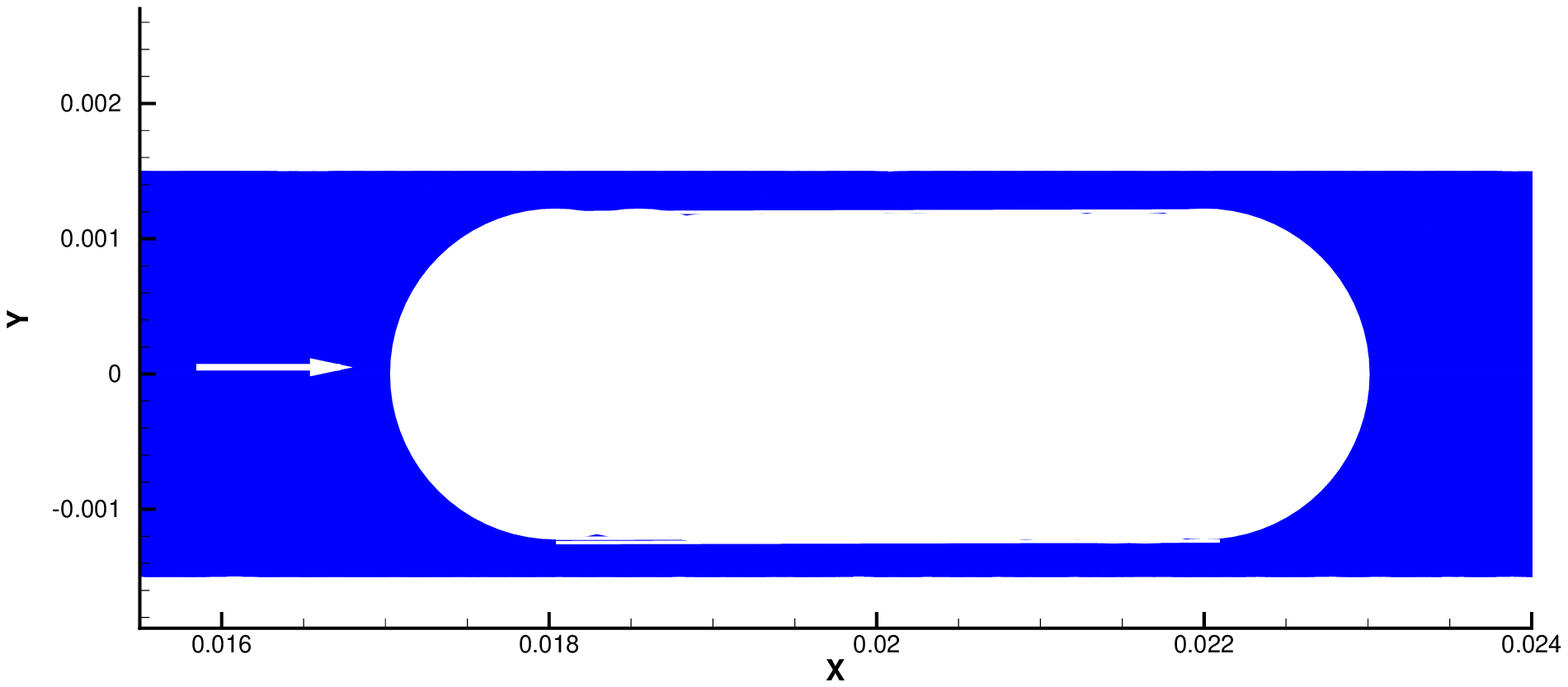}}}
\caption{Schematic of the problem. (a) a small wound with the diameter $d=2$ mm in the blood vessel ($d=3$ mm); (b) a big wound with the length 6 mm and the diameter 2 mm. The blood vessel diameter is 3 mm.}
\label{fig:wound_geometry}
\end{figure}

In the present paper the vessel with a diameter of 3 mm is considered. The fully developed
velocity profile is prescribed at the inlet of blood vessel, while
zero gradient velocity boundary condition is applied on the outlet plane. The
pressure at the wound is equal to the tissue pressure. At vessel
inlet, the blood flow cross-sectional average velocities are set
at 0.016 m/s and 0.13 m/s. These imposed velocities correspond to the velocities in
veins and arteries with the diameter of 3 mm \cite{Hand1998Ch8}.
Two wound shapes, namely, the small circular wound with a diameter of
2 mm and a big wound of 6 mm in length and 2 mm in diameter will
be investigated (Fig. \ref{fig:wound_geometry}).


The single element HIFU transducer used in this study is
spherically focused with an aperture of 12 cm and a focal length
of 12 cm. In this study, the transducer with the frequency $f_0 = 1.0$ MHz is considered. Focal intensity is 2240 W/cm$^2$,
and the sonication time is 0.6 second.
 The parameters used in the current simulation are listed
in Table \ref{tab:coefficient} \cite{Duck1990}.

\begin{table}[h]
  \centering
   \caption{Acoustic and thermal properties for the tissue and blood.}
   \label{tab:coefficient}
 \begin{tabular}{|c|c|c|c|c|c|}
    \hline
    {\bf Tissue} & {\bf ~$c_0$~$(\frac{m}{s})$} & {\bf ~$\rho$~$(\frac{kg}{m^{3}})$} & {\bf ~$c$~$(\frac{J}{kgK})$} & {\bf ~$k$~$(\frac{W}{mK})$} & {\bf ~$\alpha$~$(\frac{Np}{m})$} \\
    \hline
    tissue & $1540$ & $1055$ & $3600$ & $0.512$ & $8.1 f$\\
    \hline
    Blood & $1540$ & $1060$ & $3770$ & $0.53$ & $1.5 f$\\
    \hline
  \end{tabular}
\end{table}


\section{Results and discussion} \label{sec:results}
\noindent

As it was already mentioned in introduction, heating is considered as one of the main consequencies during acoustic hemostasis. Increase of the temperature above a certain value allows cauterizing of the bleeding site. The temperature of 70$^0$C can be assumed as the threshold temperature for the acoustic hemostasis \cite{VaezyMartin2001}.  However, blood flow out of the wound may carry away the heat at the bleeding site and significantly reduce the temperature at the bleeding site. Blood also has a lower absorption than a tissue. For example, in Table 1 it can be seen that absorption coefficients of liver tissue and blood differ by about five times. Therefore if blood is still flowing out of the wound it is quite difficult to seal the wound. First of all, it is necessary to reduce or stop the flow out of the wound, then thermal energy can be applied to cauterize the bleeding site and seal the wound.
In the experiments \cite{VaezyMartinUMB1998} with the wound exposed to air it was shown that blood flow out of the wound can be stopped after applying focused ultrasound energy to the bleeding site. Focused ultrasound induces streaming of the blood away from the focus of the transducer. This effect is called acoustic streaming.  Thus, physical processes involved in acoustic hemostasis can be described in the following way: first, acoustic streaming effect reduces or stops the flow out of the wound, afterwards thermal effect is used to seal the wound. In the following sections we are going to investigate numerically the acoustic streaming effect for two wound shapes and different sonication angles and thermal effect for different sonication angles.

\subsection{Importance of acoustic streaming}

\subsubsection{Bleeding in a small wound}

In the previous studies \cite{SolovchukSheuJASA,SolovchukSheuATE2013} it was shown that focused ultrasound can induce acoustic streaming velocities up to 100 cm/s in the blood vessel and can affect the ultrasound heating. When blood vessel was placed perpendicularly to the acoustic axis, acoustic streaming velocity magnitude becomes smaller comparing with that of the parallel blood vessel orientation.

When the focus of HIFU transducer is directed towards a bleeding site, the local absorption of acoustic
energy supplies an extra momentum to the fluid and this force can result in streaming of the blood in the direction away from the
focus of transducer. Usually in the acoustic hemostasis experiments blood vessel was located perpendicularly to the acoustic axis. However it is not very clear what blood vessel orientation is the optimal one. For the case of big wound different focal point locations and scanning path planning can be chosen differently.
So we are going to investigate the effects of blood vessel orientation (sonication angle) and focal point location.

Let's consider  a  hole (wound) on the blood vessel wall. The diameter of this hole is 2 mm. The diameter of
the blood vessel is 3 mm,  and the  maximum velocities are 3.2 cm/s (vein) and 26 cm/s (artery). Acoustic streaming velocity magnitude is 30 cm/s (without acoustic  streaming  the maximum velocity
in the vein is 3 cm/s). Acoustic streaming velocity magnitude is one  order of magnitude larger than the blood flow
velocity in the vein.

In Figs. 2, 3 the velocity profiles in vein and artery are presented with and without incident focused ultrasound. In Tables 2, 3 mass fluxes at the inlet and two outlets (outlet and wound) are presented in the vein and in the artery. Mass flux at the vessel inlet is
equal to 100\%. Without AS, 78\% of the total mass flux comes out of the wound. If we switch the
transducer, radiation force will cause the acoustic streaming flow  to occur. With the acoustic streaming effect being taken into account, the bleeding in the vein can be completely stopped. However, there is still a small bleeding out of the artery. In the artery the blood flow out of the wound can be reduced by an amount from 45 \% to 29 \%. In order to stop  blood flow out of the wound, higher power depositions should be considered.

Simulations show that acoustic streaming velocity profile reaches the steady state within a very short time interval of 0.12 s, within which the bleeding can be stopped or sufficiently reduced. This prediction is in agreement with the experimental observations \cite{VaezyMartinUMB1998}.

In the next section we will show that for different sonication angles blood flow can be stopped even in the artery.

\begin{table}[!hbp]
\centering
\caption{Acoustic streaming effect on the mass flux in the small wound in the vein ($u=1.6$ cm/s).}
\label{tab:AS_smallWound_vein}
\enspace
\begin{tabular}{cccc}
\toprule
{\bf Mass flux} & Outlet & Wound & Inlet  \\ \midrule
Without AS  & 22 \%  & 78 \% & 100 \%  \\
With AS & 100 \% & 0 & 100 \%  \\
\bottomrule
\end{tabular}
\end{table}

\begin{table}[!hbp]
\centering
\caption{Acoustic streaming effect on the mass flux in the small wound in the artery ($u=13$ cm/s).}
\label{tab:AS_smallWound_artery}
\enspace
\begin{tabular}{cccc}
\toprule
{\bf Mass flux} & Outlet & Wound & Inlet  \\ \midrule
Without AS  & 55 \%  & 45 \% & 100 \%  \\
With AS & 71 \% & 29 \% & 100 \%  \\
\bottomrule
\end{tabular}
\end{table}

\subsubsection{Bleeding in a big wound}

In Tables 4, 5 mass fluxes at the inlet and two outlets of the blood vessel with a big wound are presented for the cases with  and without considering acoustic streaming effect. The focal point is located at the center of the wound. For a larger wound it is more difficult to stop bleeding. When we take into account the acoustic streaming effect the bleeding out of the wound in the vein can be stopped. However, there is still a flow of the blood out of the wound in the artery. The blood flow out of the wound in the artery is reduced from 74 \% to 53 \% of the total mass flux due to the acoustic streaming effect. In Fig. 4 velocity profiles in the artery are presented for the cases with and without focused ultrasound.
It can be seen that the bleeding is stopped in the small focal area. This means that in the focal area we can seal the bleeding site by heating.

In Fig. 5 mass flux out of the wound is presented as the function of focal point location along the axis of the big wound. When the focal point is located at the rear of the wound, the largest mass flux and consequently the largest bleeding occur. When the focal point is located close to the front of the wound ($x=0.0175$), the mass flux is minimal. Therefore to reduce the bleeding in the large wound, focal point should be located at the front of the wound.
In Fig. 6 mass flux out of the wound as the function of time is presented at different focal point locations. Within 0.12 s blood flow becomes steady. The smallest mass flux occurs at the condition when the focal point is located at the front of the wound. The worst case happens when the focal point is located at the rear of the wound.

Several ways of ultrasound sonications can be applied to stop
bleeding in the wound. In the work of \cite{VaezyMartinJVS1999} mechanical scanning of HIFU probe was used to stop bleeding in a punctured artery. The frequency of the scanning was 15 or 25 Hz, the amplitude of the scanning was equal to the length of the wound (5-10 mm). In Fig. 7 the predicted evolution of mass flux for the big wound is presented for the case of mechanical scanning of HIFU beam along the wound. Focal point was oscillating around the center of the wound with  the frequency 25 Hz. We can see that mass flux value is oscillating around the value of mass flux for the central location of the focal point. Decrease of the oscillating frequency will lead to the increased oscillating amplitude. The mass flux is very close to that of the case when the focal point is at the center of the wound.

\begin{table}[!hbp]
\centering
\caption{The effect of acoustic streaming on the mass flux in the big wound in the vein ($u=1.6$ cm/s).}
\label{tab:AS_bigWound_vein}
\enspace
\begin{tabular}{cccc}
\toprule
{\bf Mass flux} & Outlet & Wound & Inlet  \\ \midrule
Without AS  & {\textbf{3}} \%  & 97 \% & 100 \%  \\
With AS & 100 \% & 0 & 100 \%  \\
\bottomrule
\end{tabular}
\end{table}

\begin{table}[!hbp]
\centering
\caption{The effect of acoustic streaming on the mass flux in the big wound in the artery ($u=13$ cm/s).}
\label{tab:AS_bigWound_artery}
\enspace
\begin{tabular}{cccc}
\toprule
{\bf Mass flux} & Outlet & Wound & Inlet  \\ \midrule
Without AS  & 26 \%  & 74 \% & 100 \%  \\
With AS & 47 \% & 53 \% & 100 \%  \\
\bottomrule
\end{tabular}
\end{table}

\subsubsection{Bleeding in a big wound. Different sonication angles.}

In Fig. 8 the evolution of the predicted mass flow out of the big wound is presented for different sonication angles in the artery. Focal point is located at the center of the wound. The best sonication angles are from 0 to 90 $^0$. At the angle 0$^0$ it's possible to completely stop bleeding within 0.1 s. Because ultrasound beam is quite narrow, for 0$^0$ sonication it's quite difficult to locate the ultrasound beam precisely on the wound. For the sonication angle 45$^0$ there is only small bleeding at the rear of the wound (Fig. 9). In the experiments, 90$^0$ sonications were mostly used. From the simulation point of view, it is better to use 45$^0$.  The predicted velocity magnitude in the artery with the big wound for two different sonication angles 45 $^0$ and 135 $^0$ are presented in Fig. 9. For 135$^0$ sonication there is a reverse flow near the center of the wound. This reverse flow acts like a dam and therefore velocity magnitude in the beginning of the wound is increased. Comparison of the mass fluxes with and without AS for this sonication angle shows that 135$^0$ sonication increases bleeding and should be avoided in the treatment.

In Fig. 10 the evolution of the predicted mass flow out of the big wound is presented for the two sonication angles 45$^0$ and 135$^0$ and two locations of the focal point: at a location in front of the wound and at the rear of the wound.  The corresponding velocity magnitude for the sonication angle 45$^0$ can be seen in Fig. 11. When the focal point is located at the rear of the wound and the angle is 45$^0$, the bleeding can be stopped. When the focal point is located in front of the wound and the angle is 45$^0$, there is still a small blood flow out of the wound at the rear of the wound. For 135$^0$ an opposite result is obtained. For 90$^0$ and 135$^0$ sonications focal point should be located at the front of the wound. The numerical simulations show that 135$^0$ sonication should be avoided, because in this case the degree of bleeding increases. The smallest mass flux occurs at 45$^0$ sonication when the focal point is located at the rear of the wound. Sonication angles between 45 and 90 degrees should be considered in order to reduce the flow out of the wound. In some cases flow out of the wound can be stopped.


Although in the experiments 90$^0$ sonications are mostly used \cite{VaezyMartinUMB1998}, we have shown that for a big wound the optimal focal point location is at the rear of the wound and the optimal angle is 45$^0$. In this case the flow out of the wound can be completely stopped. However the wound is not sealed yet, we should apply thermal energy in order to cauterize the bleeding site. In the following section we are going to investigate the thermal effect of ultrasound for different sonication angles.

\subsection{Thermal effects}


Let's study the temperature distribution in the tissue and in the blood domain during acoustic hemostasis.
In most of the experimental studies ultrasound beam was located perpendicularly to the blood vessel. However, the optimal angle between the blood vessel and ultrasound beam is not very clear and it is not always possible to locate ultrasound beam perpendicular to the blood vessel. We are going to investigate how different sonication angles affect the temperature elevation. We assumed that acoustic streaming effect has already stopped the flow out of the wound and the blood vessel is intact. Focal point is located on the blood vessel wall. Blood flow velocity in the vein is 1.6 cm/s. In Fig. 12 the predicted temperature contours at $t=0.6$ s at the cutting plane $y=0$ are presented for three different sonication angles. We can see a very small temperature increase inside the blood vessel.

The predicted acoustic streaming velocity at the cutting plane in the blood vessel without an externally applied flow is presented in Fig. 13 for three different sonication angles.
The acoustic streaming velocity magnitudes are 34 cm/s, 36 cm/s and 60 cm/s for the sonication angles $90^0$, $45^0$ (or $135^0$) and $0^0$(or $180^0$), respectively. The acoustic streaming velocity has the smallest value for $90^0$ sonication and the largest value for $0^0$ sonication. For all sonication angles, acoustic streaming velocity magnitude is up to an order of magnitude larger than velocity in blood vessel (1.6 cm/s in vein and 13 cm/s in artery). Therefore cooling due to acoustic streaming effect can prevail over the blood flow cooling and can represent the main cooling mechanism.

In Fig. 14 the predicted temperature  is presented at the focal point as the function of time for different sonication angles. It can be seen that for $90^0$ sonication the predicted temperature has the largest value and for $0^0$ sonication the predicted temperature has the smallest value. For $45^0$ and $135^0$ sonications the predicted temperatures are almost the same (about 1$^0C$ difference), because acoustic streaming velocity magnitude (35 cm/s) is an order of magnitude larger than the velocity in blood vessel (1.6 cm/s) and acoustic streaming is the main cooling mechanism in this case. The temperature around 70$^0$C on the blood vessel wall can be reached at $t=0.6$ s for $90^0$ sonication, for other sonication angles it will take a longer time to reach the temperature 70$^0$C. This shows the possibility to stop bleeding theoretically. For a smaller blood vessel the effects of blood flow
cooling and acoustic streaming will be smaller.




In the current subsection we assumed that there is no blood flow coming
out of the wound. In this case the temperature around 70$^0$C can
be achieved quite rapidly ($t<1$ s) and the wound can be sealed in a
short time. Simulation shows that $90^0$ sonication should be chosen in order to optimize the treatment.

\section*{Conclusions}
\label{Conclusions}

The mathematical model for the simulation of acoustic
hemostasis is proposed in the current paper. Our analysis is based on
the nonlinear Westervelt equation with the relaxation effect being
taken into account and the bioheat equations are applied in blood vessel
and tissue domains. The nonlinear hemodynamic equation is also
considered with the acoustic streaming effect being taken into
account.

Both thermal and  acoustic streaming effects have been
investigated in the current paper.  The importance of acoustic
streaming was examined for different blood vessel orientations and
focal point locations. Acoustic streaming velocity magnitude is up to 60 cm/s and this magnitude is several times larger than the velocity in blood vessel. If focused ultrasound beam is applied directly to the bleeding site, the
 flow out of a wound is considerably reduced due to acoustic
 streaming. Bleeding can be even completely stopped depending on the blood vessel
 orientation and the focal point location. As a result, the wound can
 be quickly sealed. Simulations show that the
temperature around 70 $^0$C can be achieved within a second on the
blood vessel wall, if there is no flow out of the wound. The
temperature inside blood vessel remains almost unchanged. Sonication angles between 45 and 90 degrees should be considered in order to reduce blood flow out of the wound. This simulation
confirms the theoretical possibility of sealing the bleeding site by
means of focused ultrasound. The blood vessel remains patent after
the treatment.

\section*{Acknowledgement}
\noindent

The authors would like to acknowledge the financial support from the Center for Advanced Study in Theoretical Sciences  and from the National Science Council of Republic of China under Contract No. NSC102-2811-M-002-125.

\clearpage

\begin{figure}
\centering
\mbox{\subfigure[]{\includegraphics[width=0.4\textwidth]{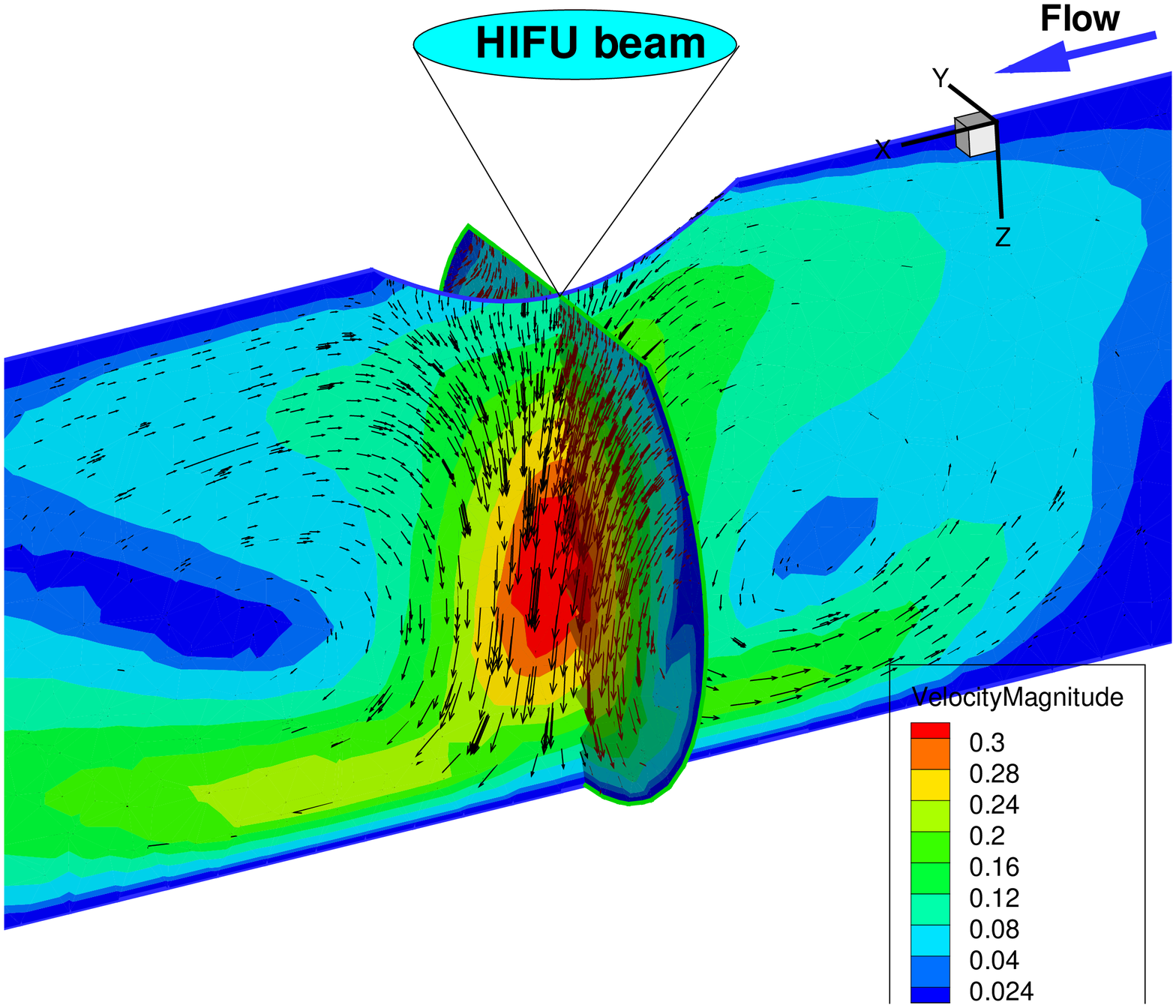}}
      \subfigure[]{\includegraphics[width=0.4\textwidth]{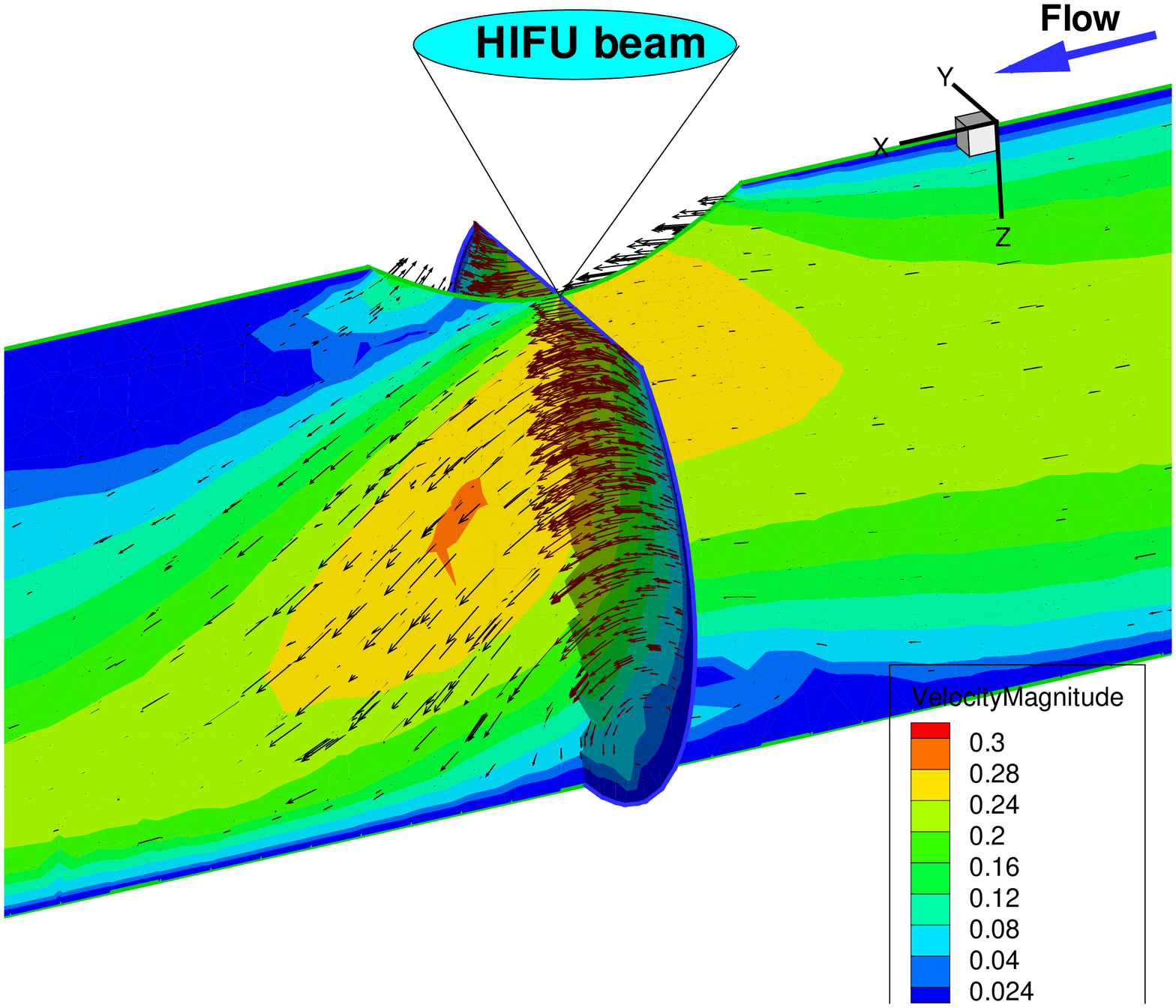}}}
\caption{The predicted velocity in vein (a) and artery (b) when the acoustic streaming effect is taken into account.}
\label{fig:smallwoundAS}
\end{figure}

\begin{figure}
\centering
\mbox{\subfigure[]{\includegraphics[width=0.4\textwidth]{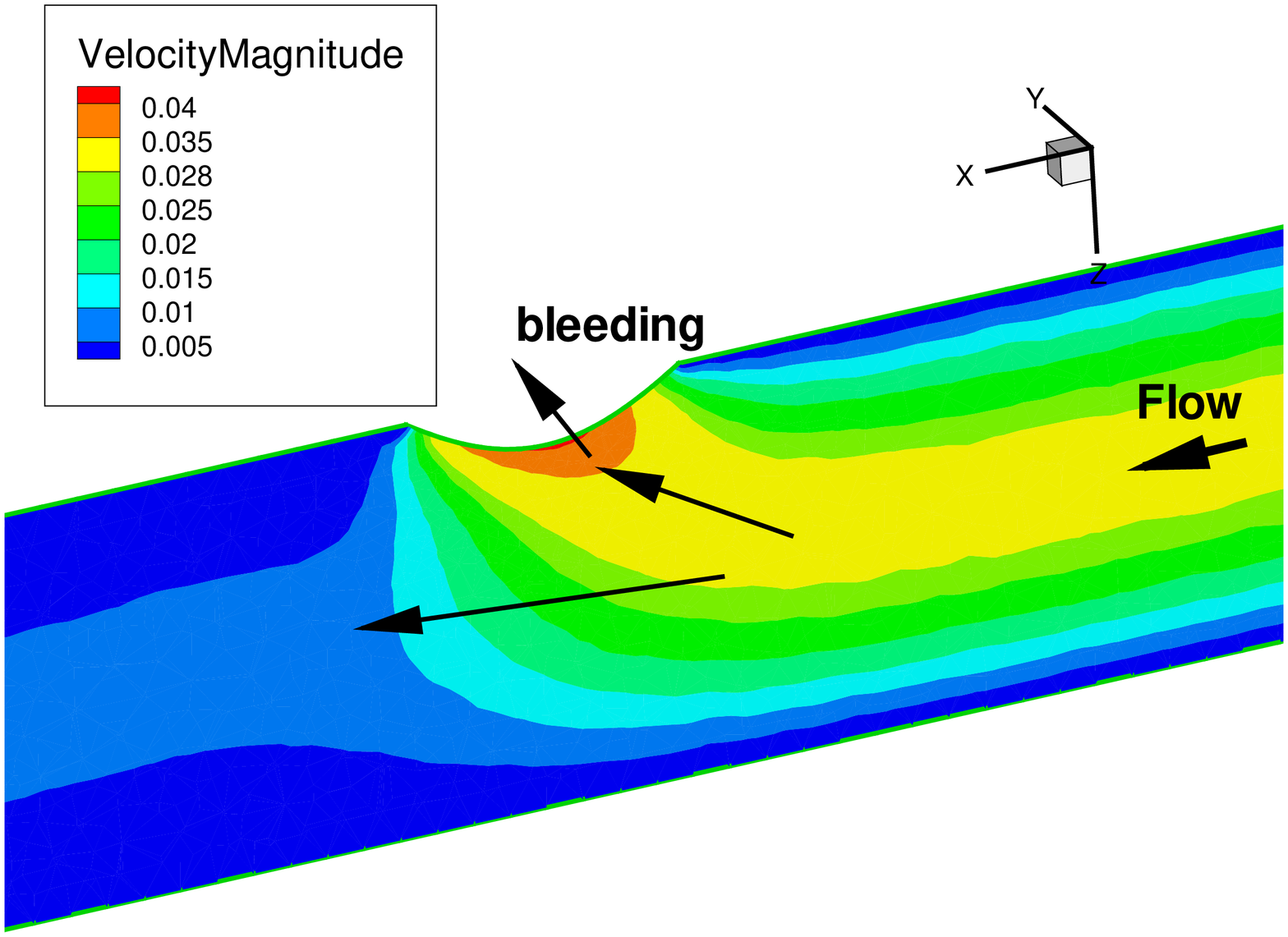}}
      \subfigure[]{\includegraphics[width=0.4\textwidth]{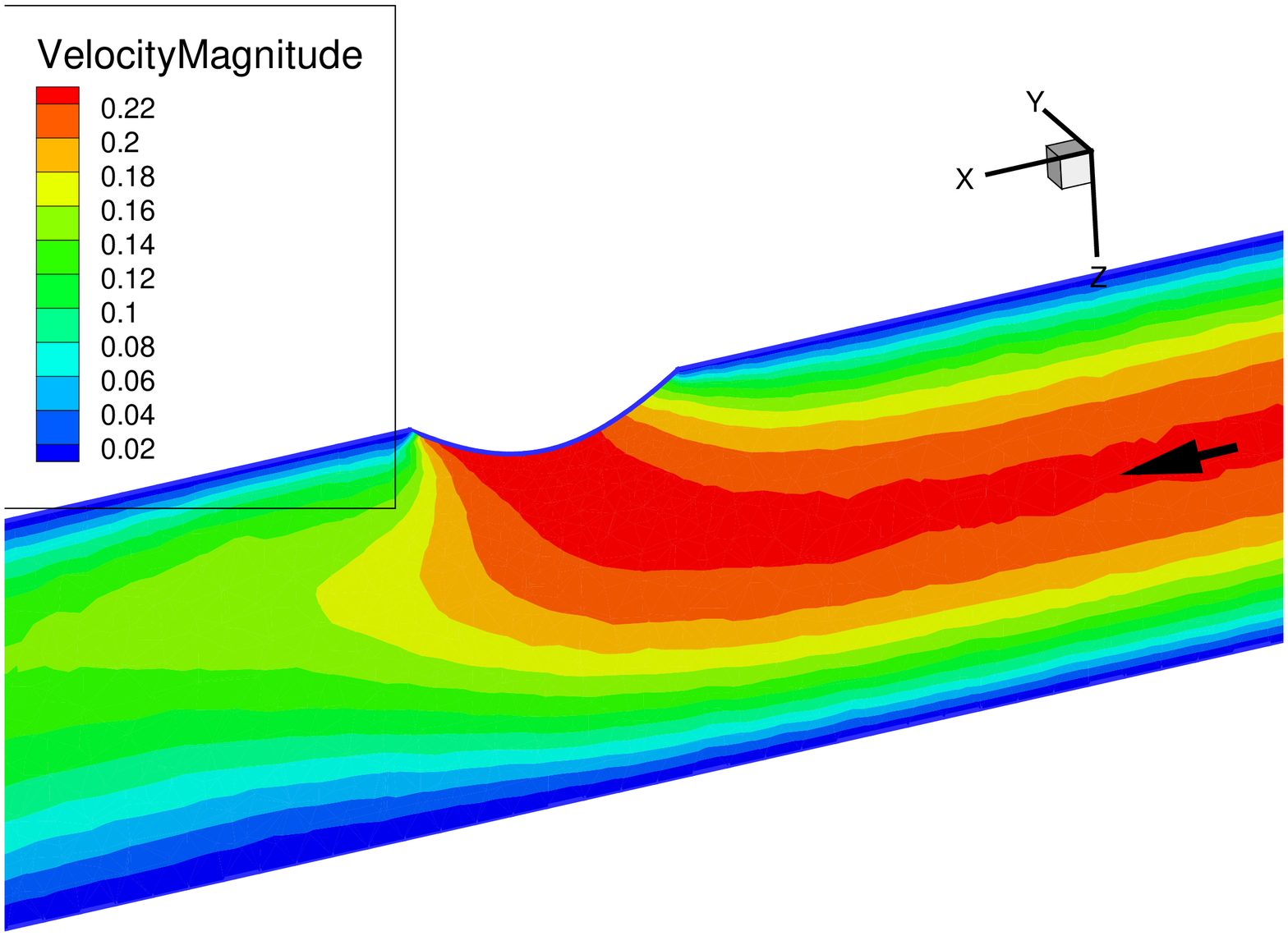}}}
\caption{The predicted velocity in vein (a) and artery (b) without acoustic streaming effect.}
\label{fig:smallwound_NoAS}
\end{figure}

\begin{figure}
\centering
\mbox{\subfigure{\includegraphics[width=0.4\textwidth]{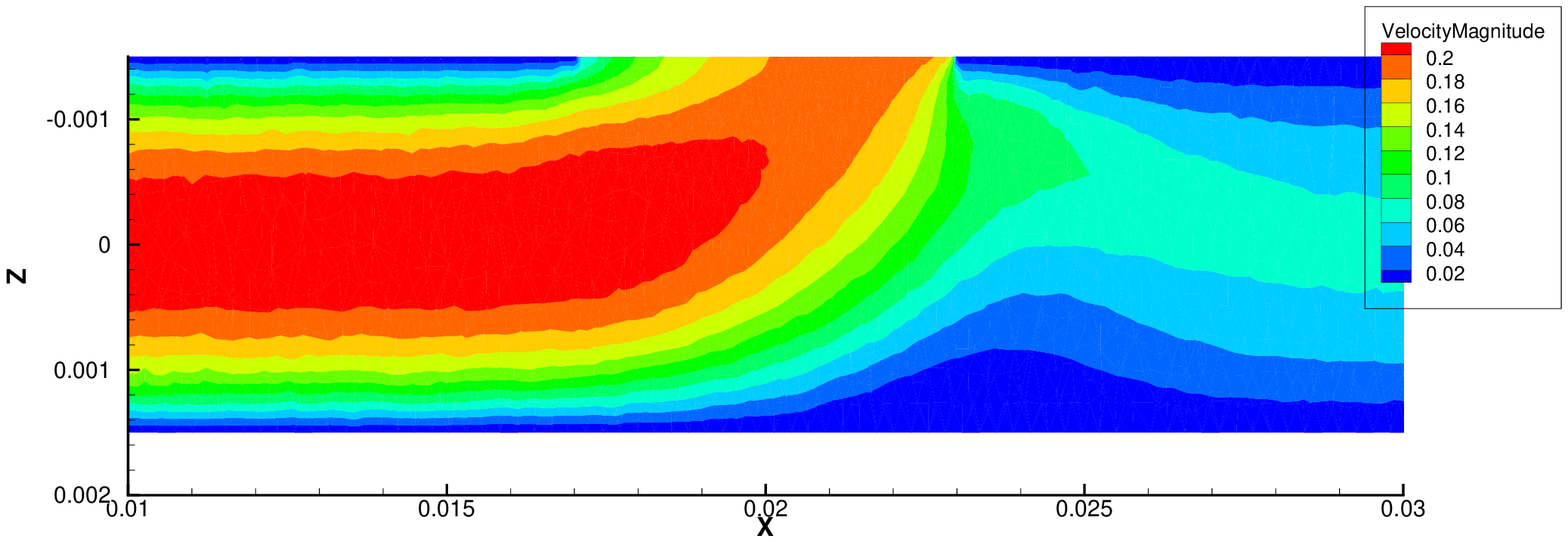}}
      \subfigure{\includegraphics[width=0.4\textwidth]{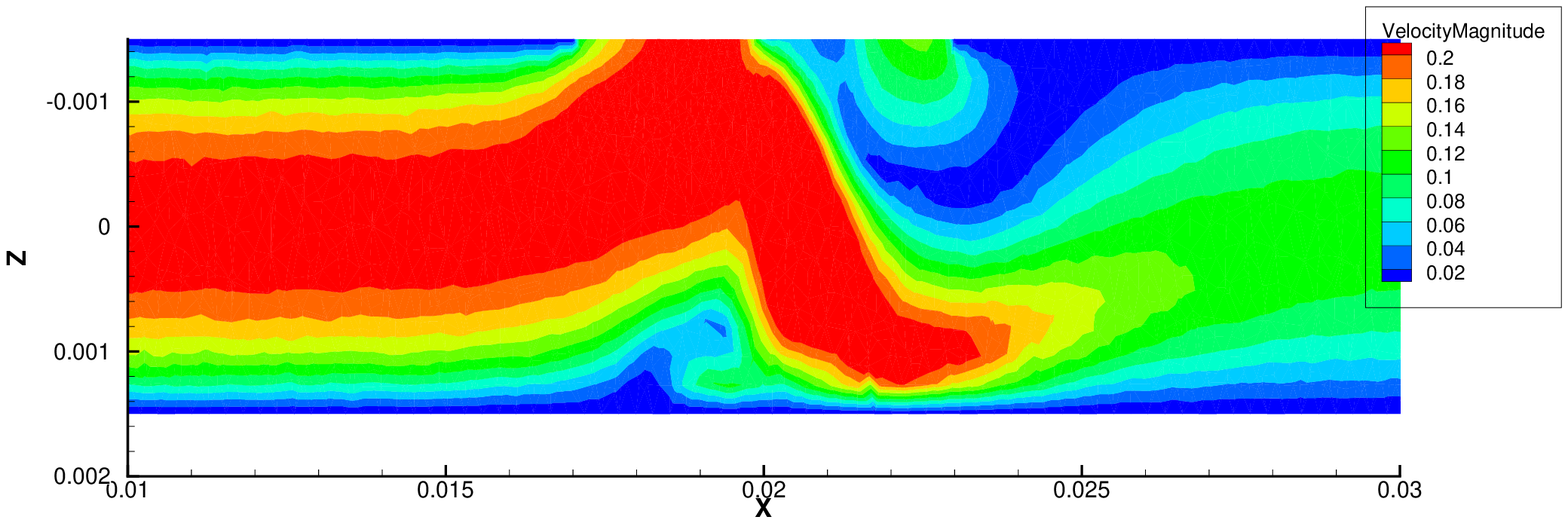}}}
\mbox{\subfigure{\includegraphics[width=0.40\textwidth]{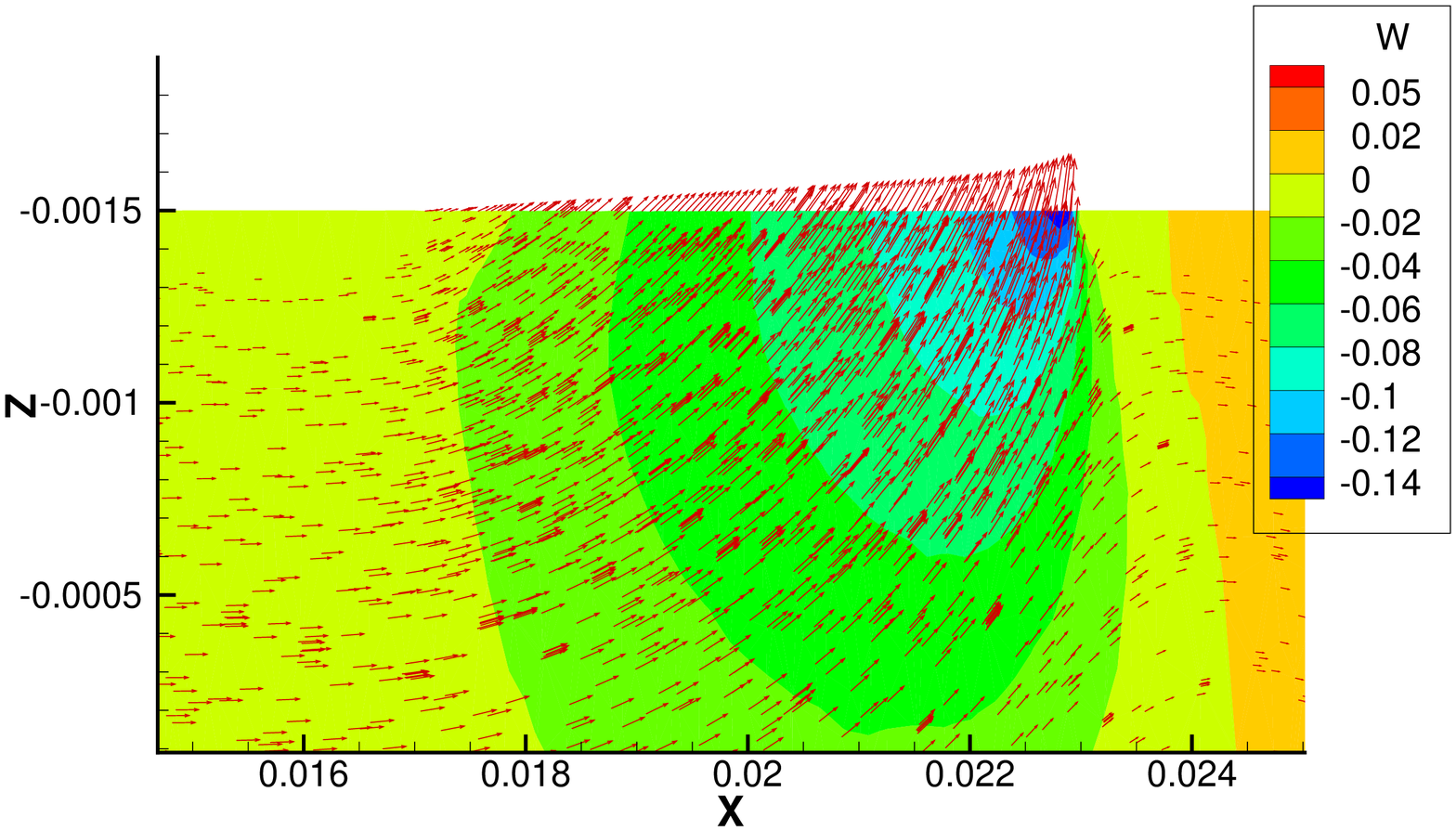}}
      \subfigure{\includegraphics[width=0.40\textwidth]{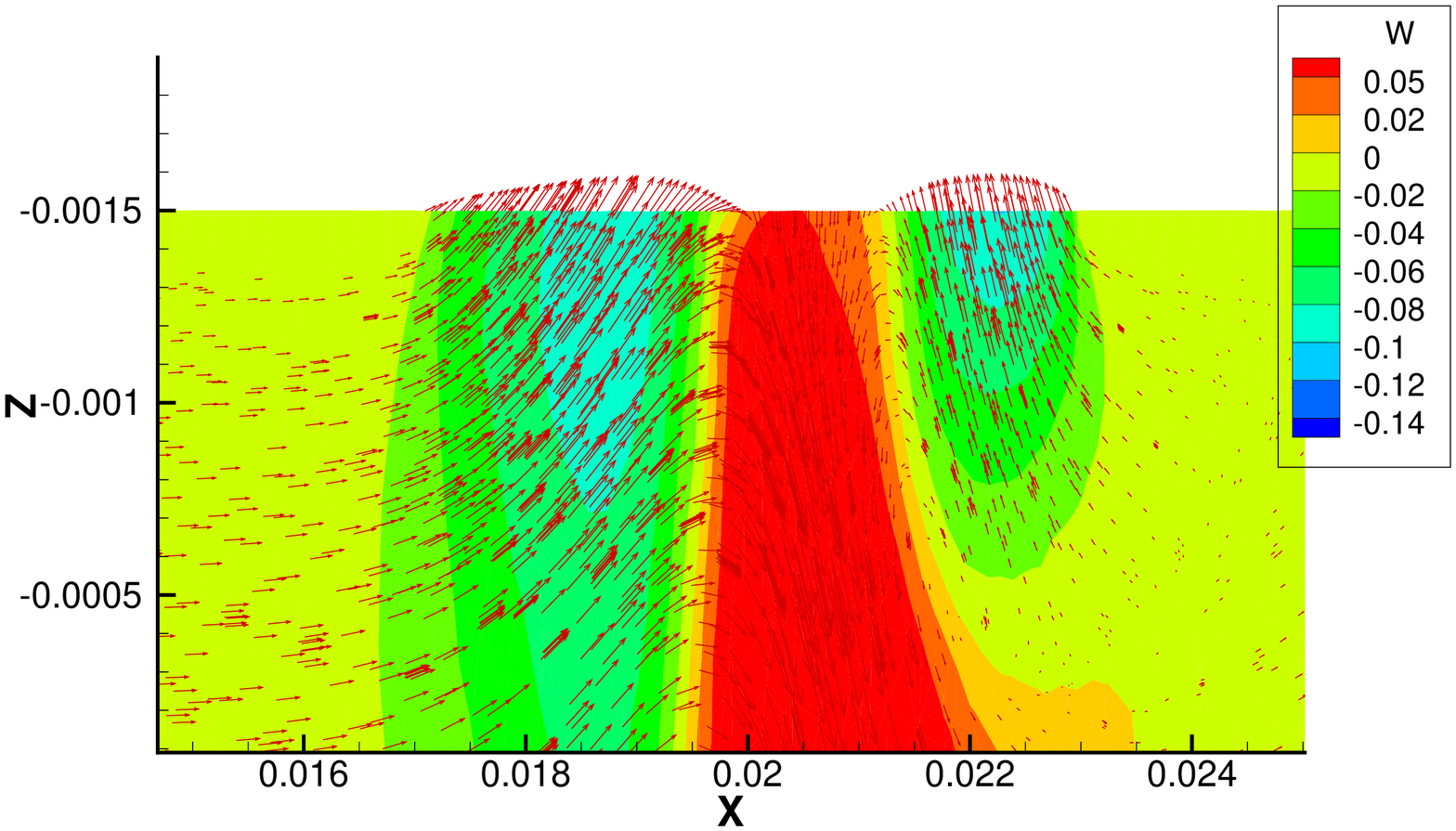}}}
\caption{The predicted velocity magnitude contours in the artery with the big wound for the cases with and without considering acoustic streaming. $w$ - velocity in $z$ direction.}
\label{fig:bigwound_center}
\end{figure}

\begin{figure}
\centering
\includegraphics[width=0.45\textwidth]{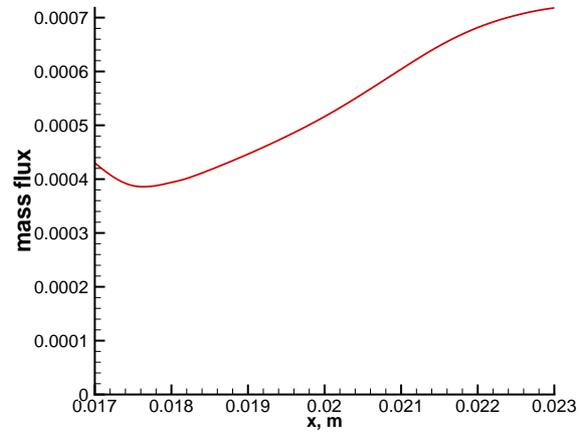}
\caption{The predicted mass flow at the big wound for the cases with different focal point locations, artery case.}
\label{fig:bigwound_Flux_vs_location}
\end{figure}

\begin{figure}
\centering
\includegraphics[width=0.45\textwidth]{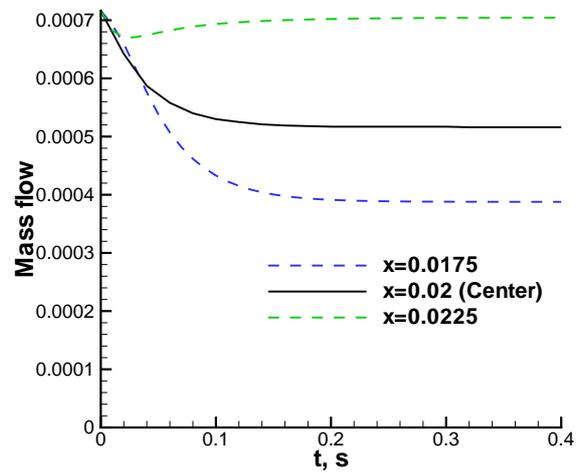}
\caption{The predicted mass flow with respect to time in the wound for the cases with different focal point locations, artery case.}
\label{fig:bigwound_perp_time_location}
\end{figure}

\begin{figure}
\centering
\includegraphics[width=0.45\textwidth]{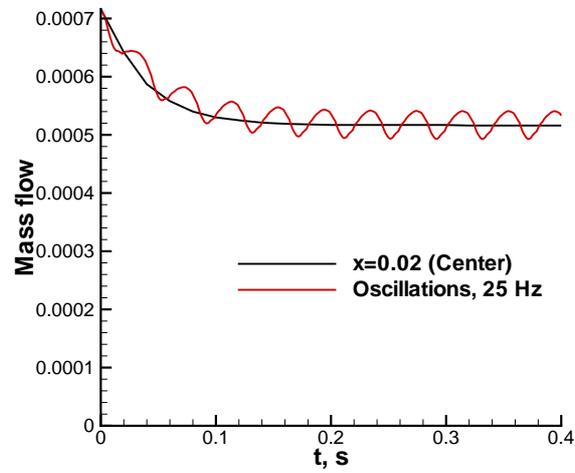}
\caption{The predicted mass flow with respect to time in the wound for two cases: 1) focal point is located at the center of the wound ($x=0.02$); 2) focal point is oscillating around the center of the wound with the frequency 25 Hz .}
\label{fig:bigwound_perp_time_oscillations}
\end{figure}

\begin{figure}
\centering
\includegraphics[width=0.45\textwidth]{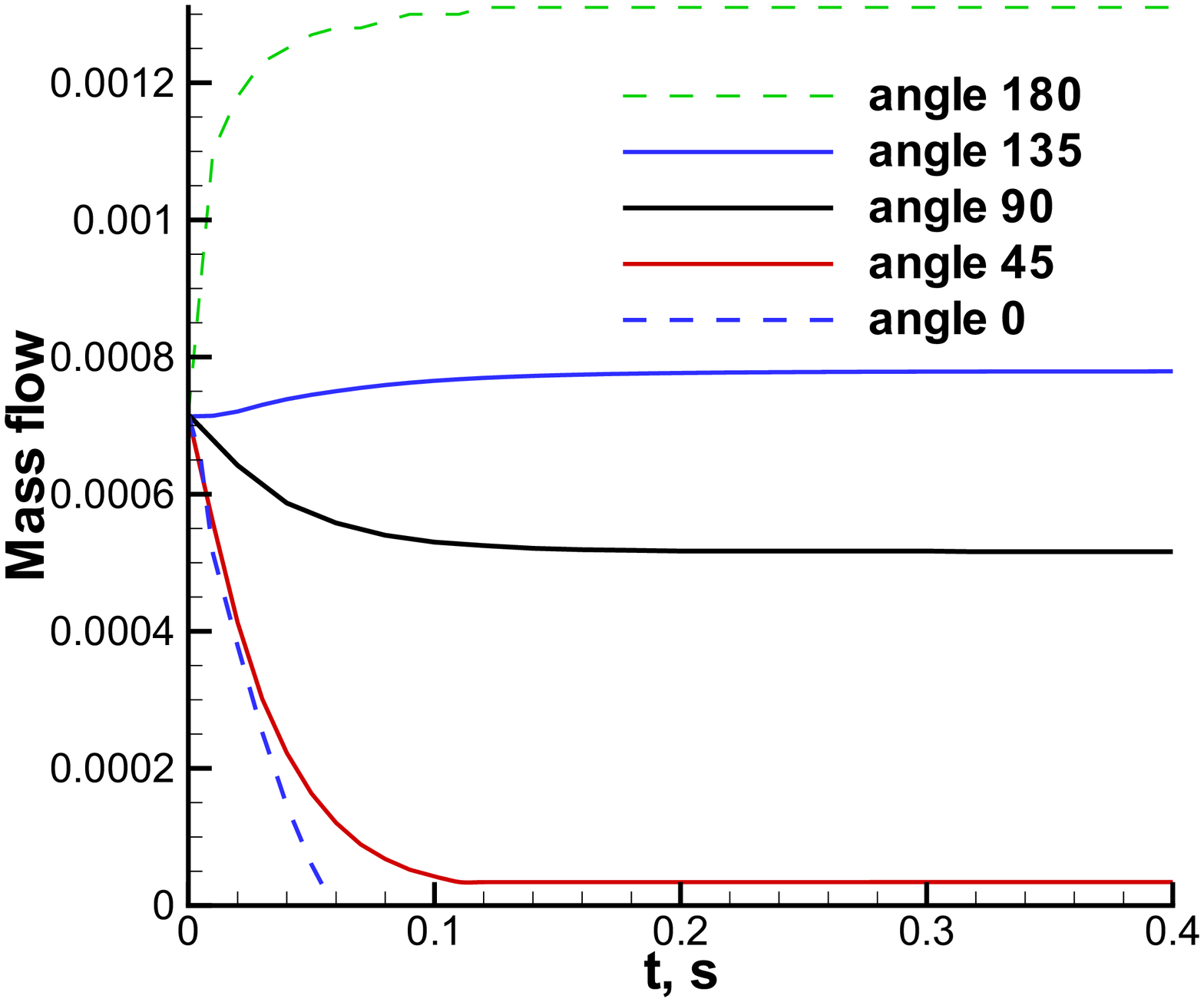}
\caption{The predicted mass fluxes with respect to time in the big wound for the cases with different sonication angles in the artery.}
\label{fig:bigwound_angle_time}
\end{figure}

\begin{figure}
\centering
\mbox{\subfigure{\includegraphics[width=0.42\textwidth]{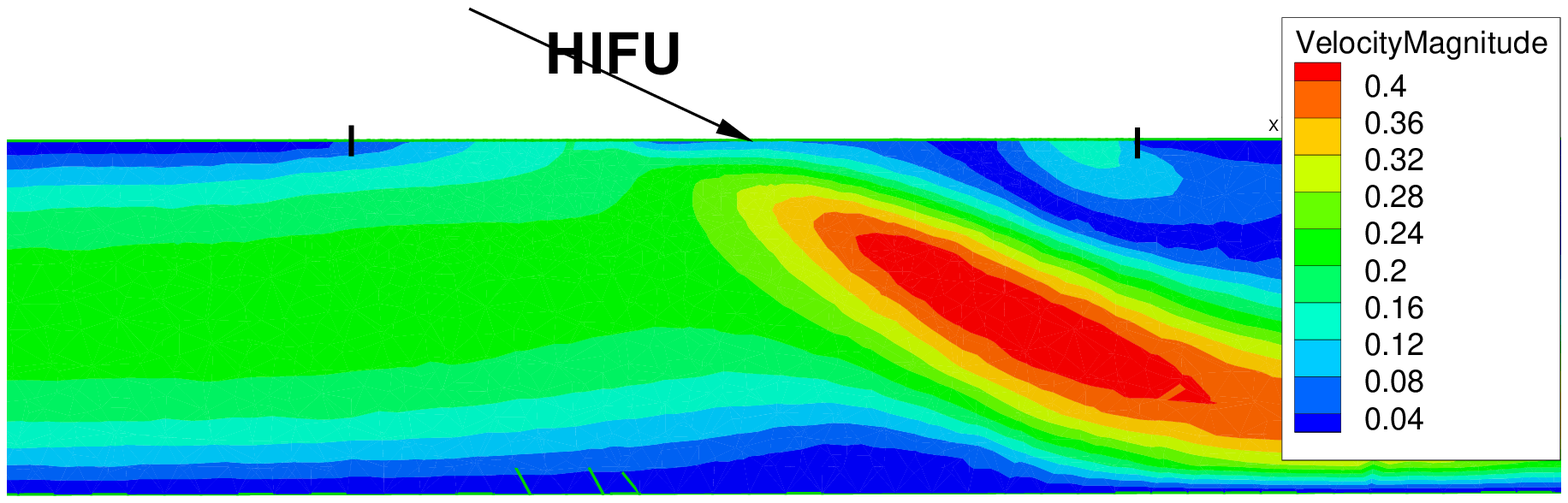}}
      \subfigure{\includegraphics[width=0.42\textwidth]{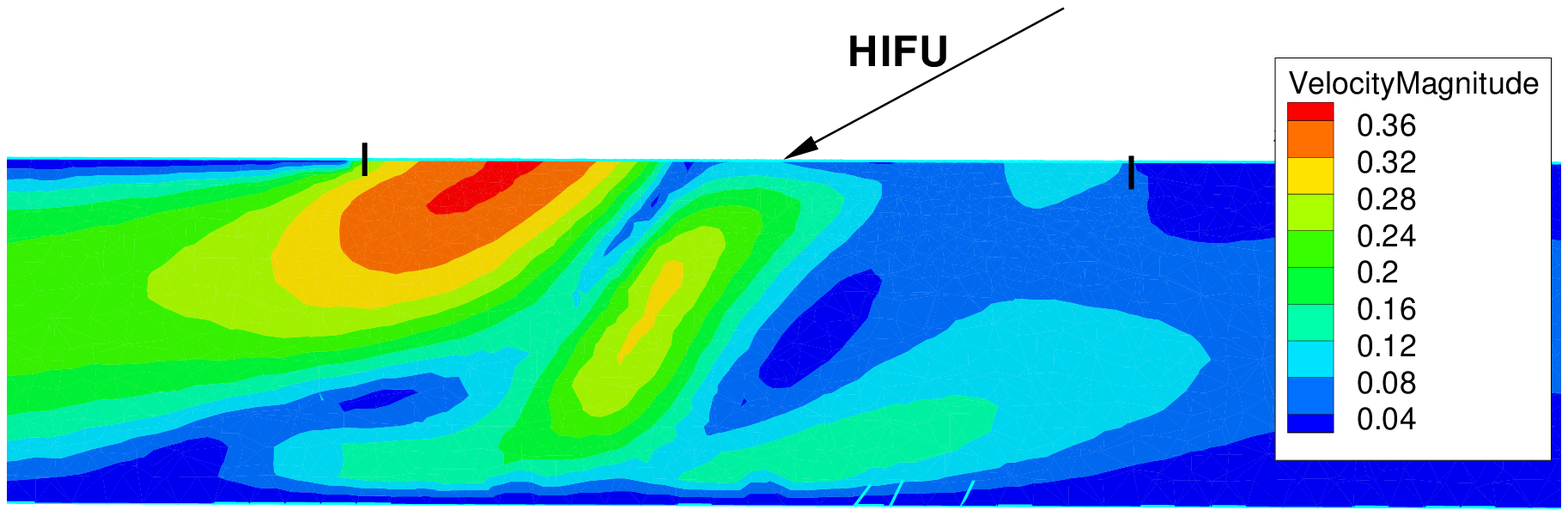}}}
\mbox{\subfigure{\includegraphics[width=0.42\textwidth]{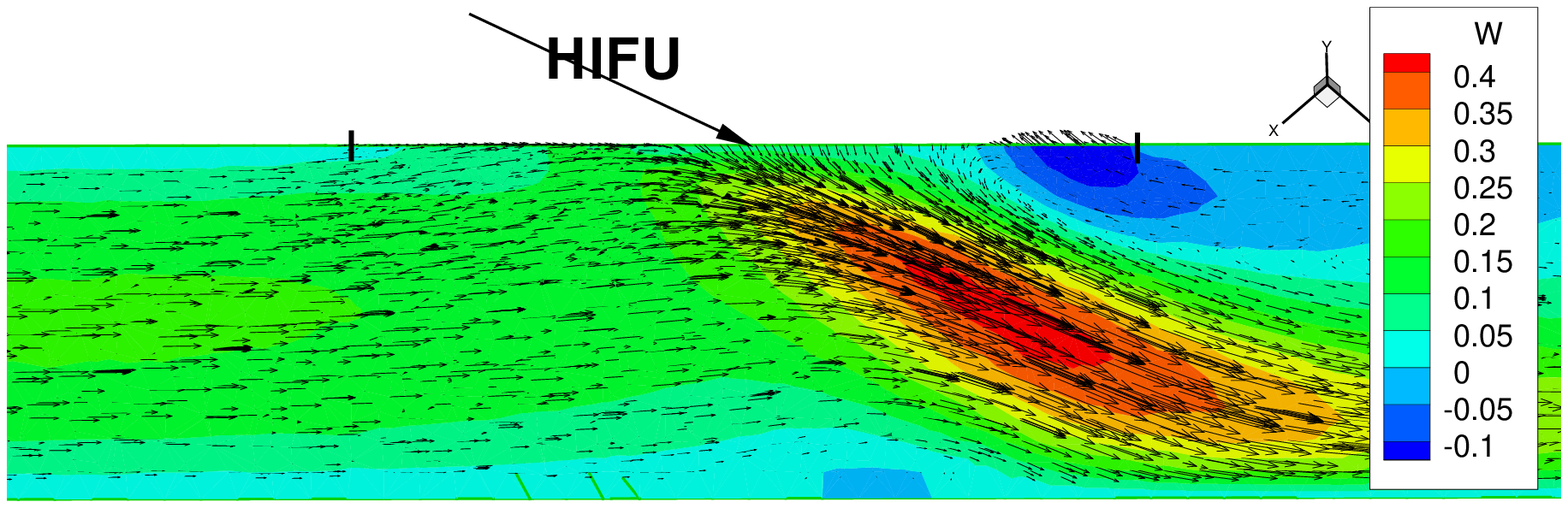}}
      \subfigure{\includegraphics[width=0.42\textwidth]{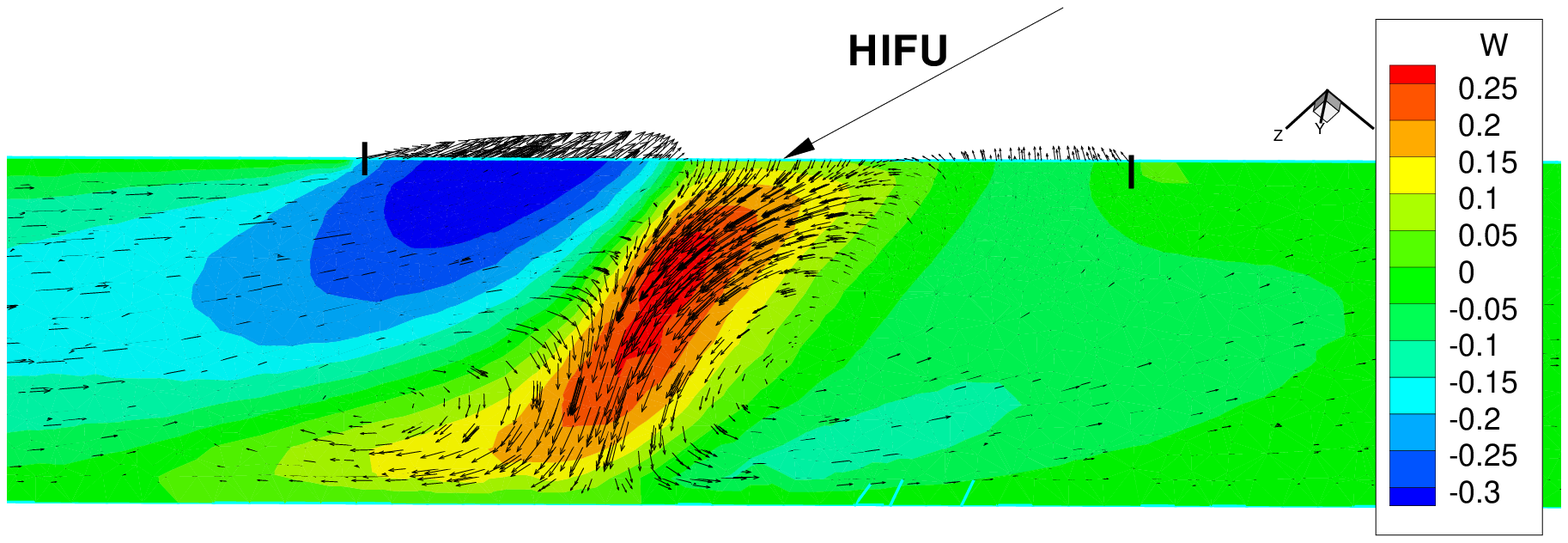}}}
\caption{The predicted velocity magnitude contours in the artery with the big wound for the sonication angles 45 $^0$ and 135 $^0$. Focal point is located at the center of the wound in the artery. $w$ - velocity in $z$ direction.}
\label{fig:bigwound_angle_velocity}
\end{figure}

\begin{figure}
\centering
\includegraphics[width=0.45\textwidth]{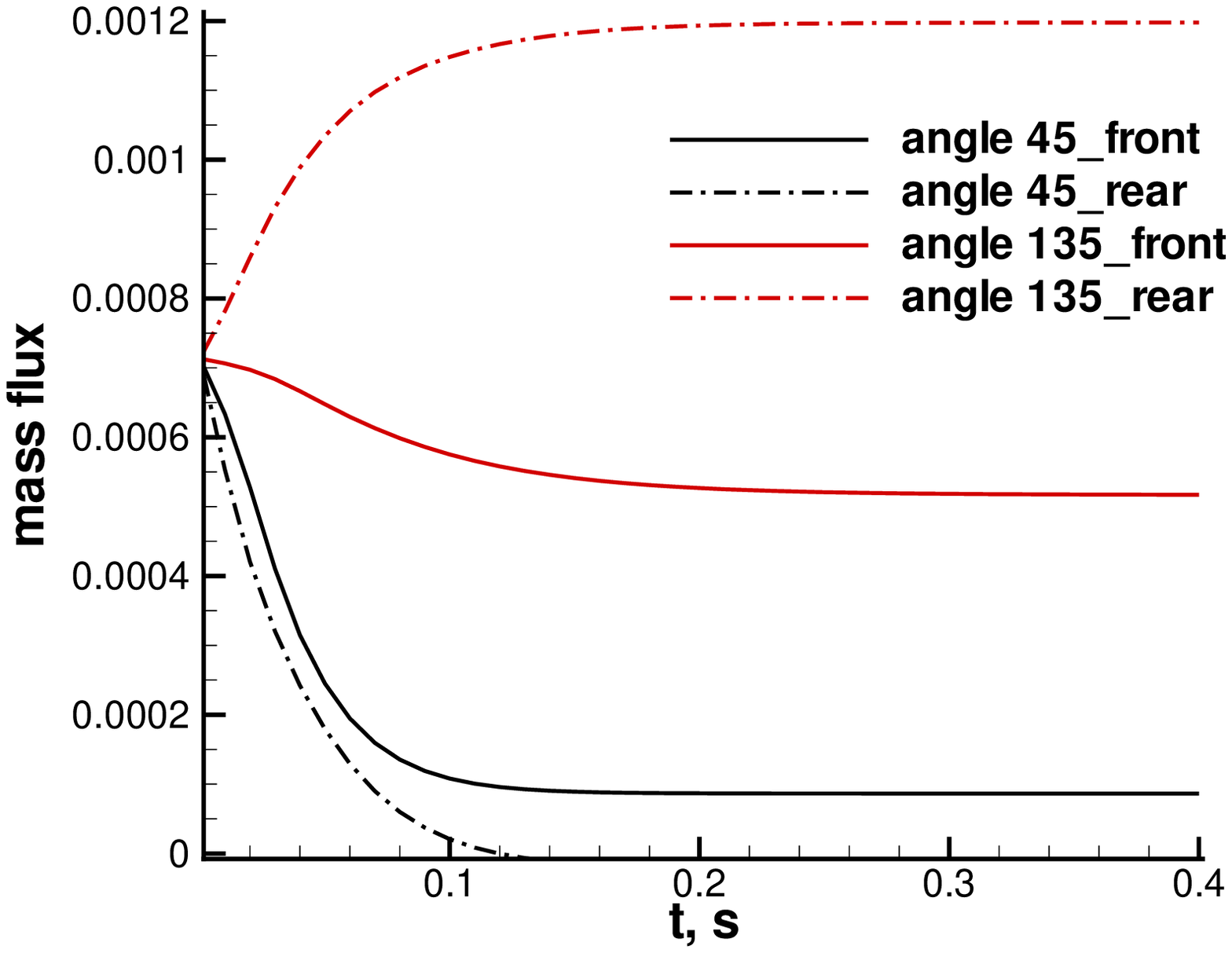}
\caption{The predicted mass fluxes with respect to time in the big wound for the cases with different sonication angles and different locations of the focal point in the artery.}
\label{fig:bigwound_angle_time_end_start}
\end{figure}

\begin{figure}
\centering
\mbox{\subfigure[]{\includegraphics[width=0.45\textwidth]{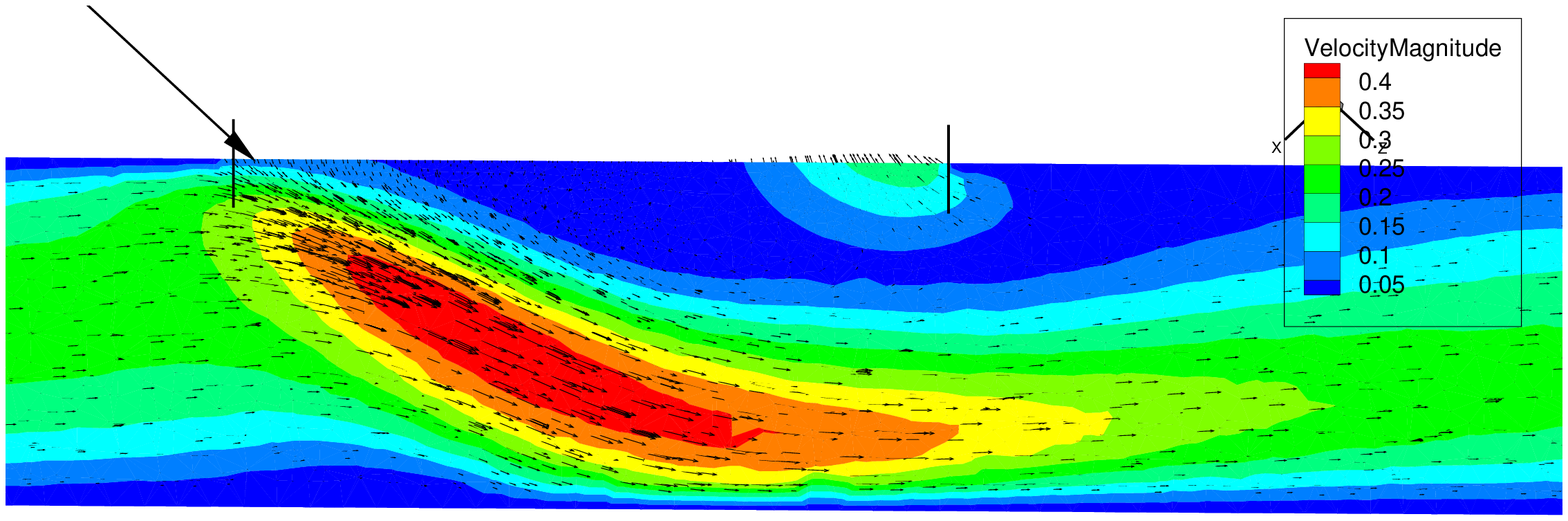}}
      \subfigure[]{\includegraphics[width=0.45\textwidth]{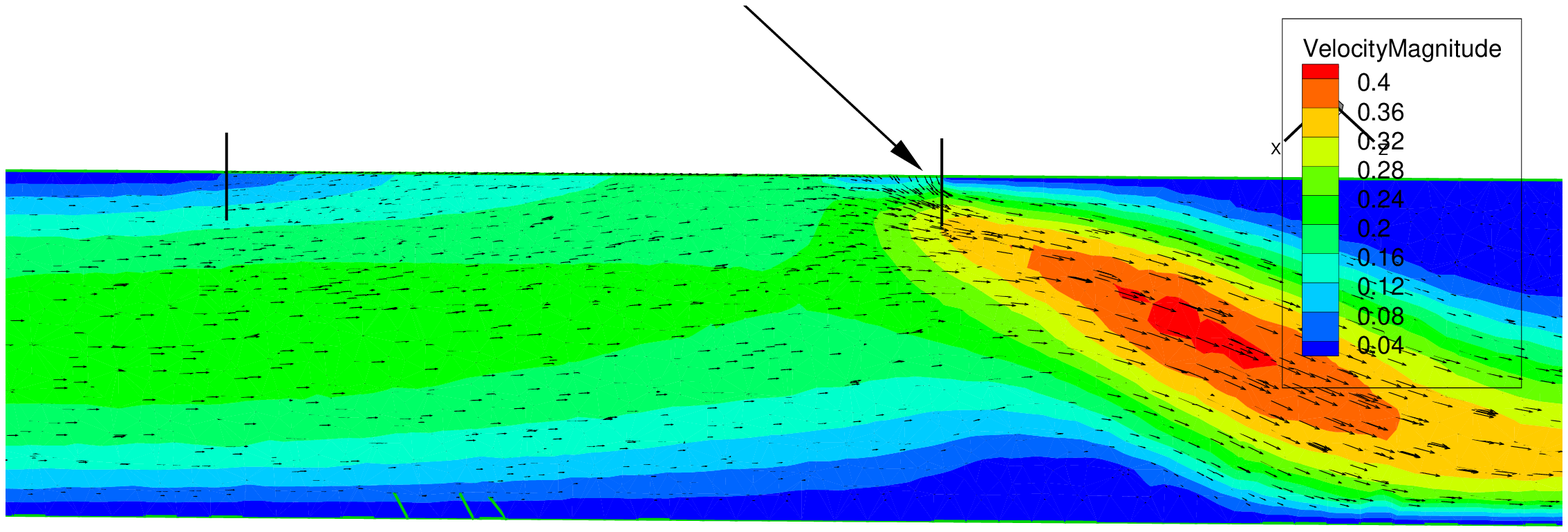}}}
\caption{The predicted velocity magnitude in the artery with the big wound for the cases with different locations of the focal point: (a) in front of the wound and (b) at the rear of the wound. The sonication angle is 45 $^0$.}
\label{fig:bigwound_45_start_end_velocity}
\end{figure}

\begin{figure}
\centering
\mbox{\subfigure[]{\includegraphics[width=0.3\textwidth]{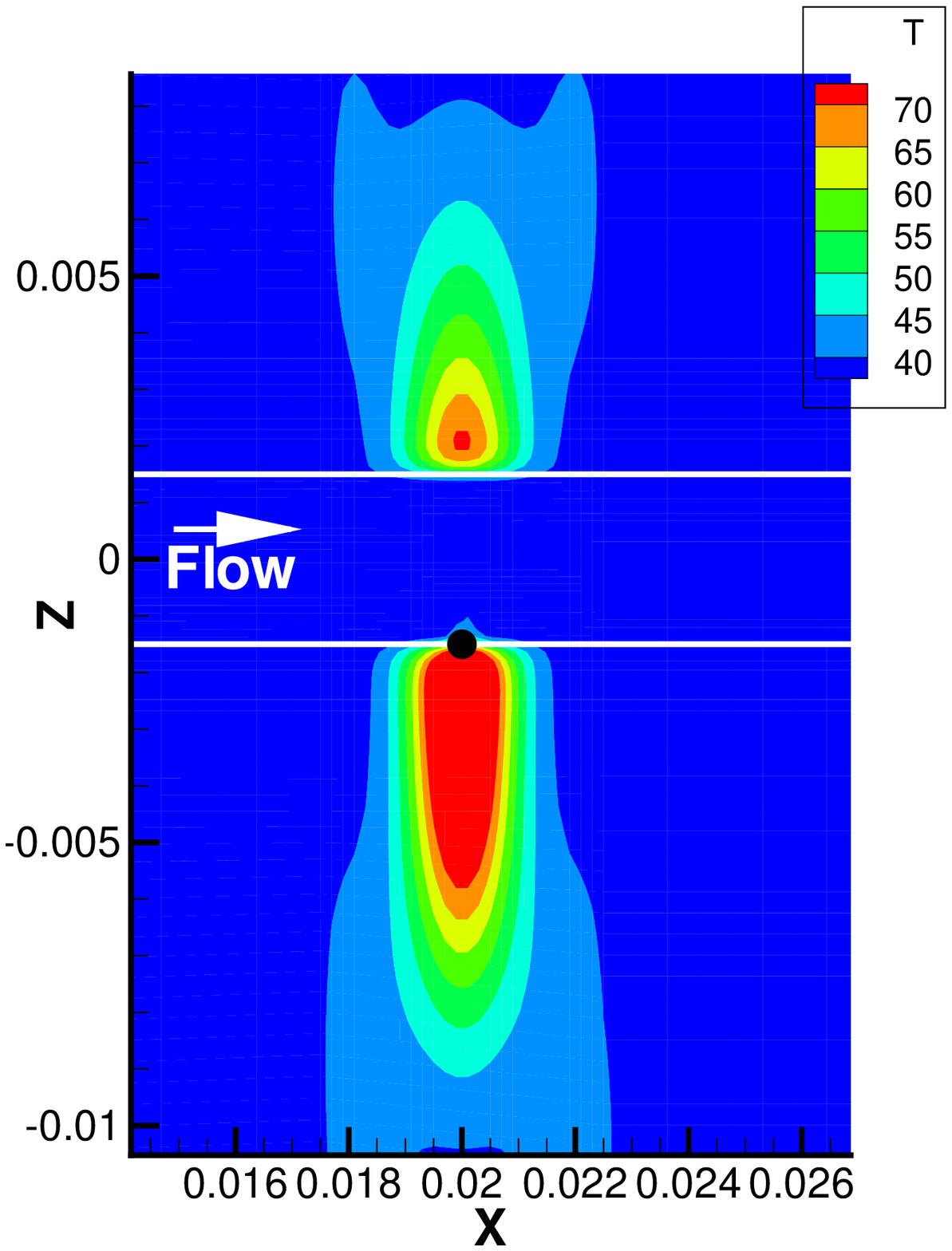}}
      \subfigure[]{\includegraphics[width=0.3\textwidth]{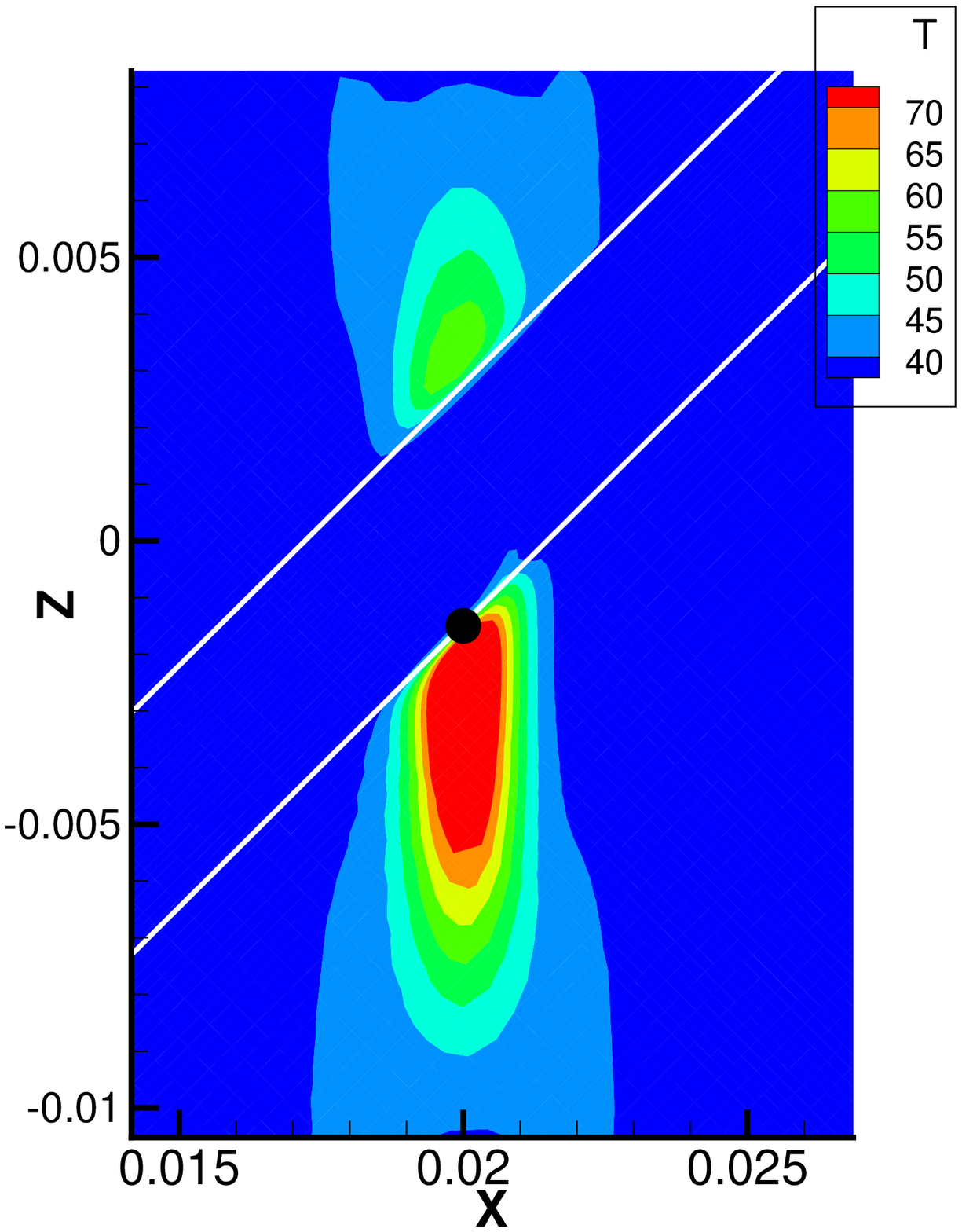}}
      \subfigure[]{\includegraphics[width=0.3\textwidth]{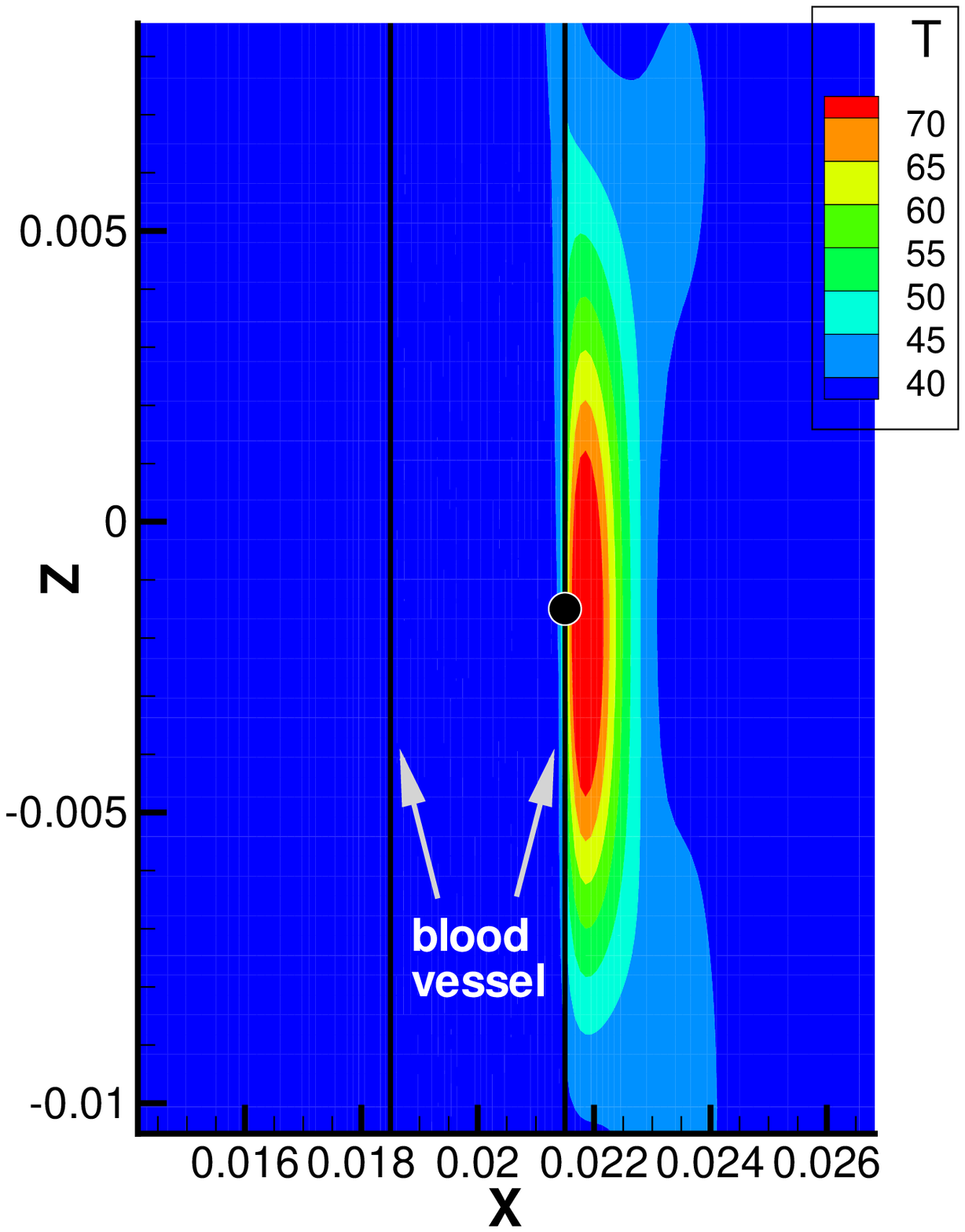}}}
\caption{The predicted temperature contours at $t=0.6$ s at the cutting plane $y=0$ for different sonication angles $90^0$ (a), $45^0$ (b) and $0^0$ (c). Focal point ($\bullet$) is located on the blood vessel wall.}
\label{fig:T_d3_t06_45_90}
\end{figure}

\begin{figure}
\centering
\mbox{\subfigure[]{\includegraphics[width=0.34\textwidth]{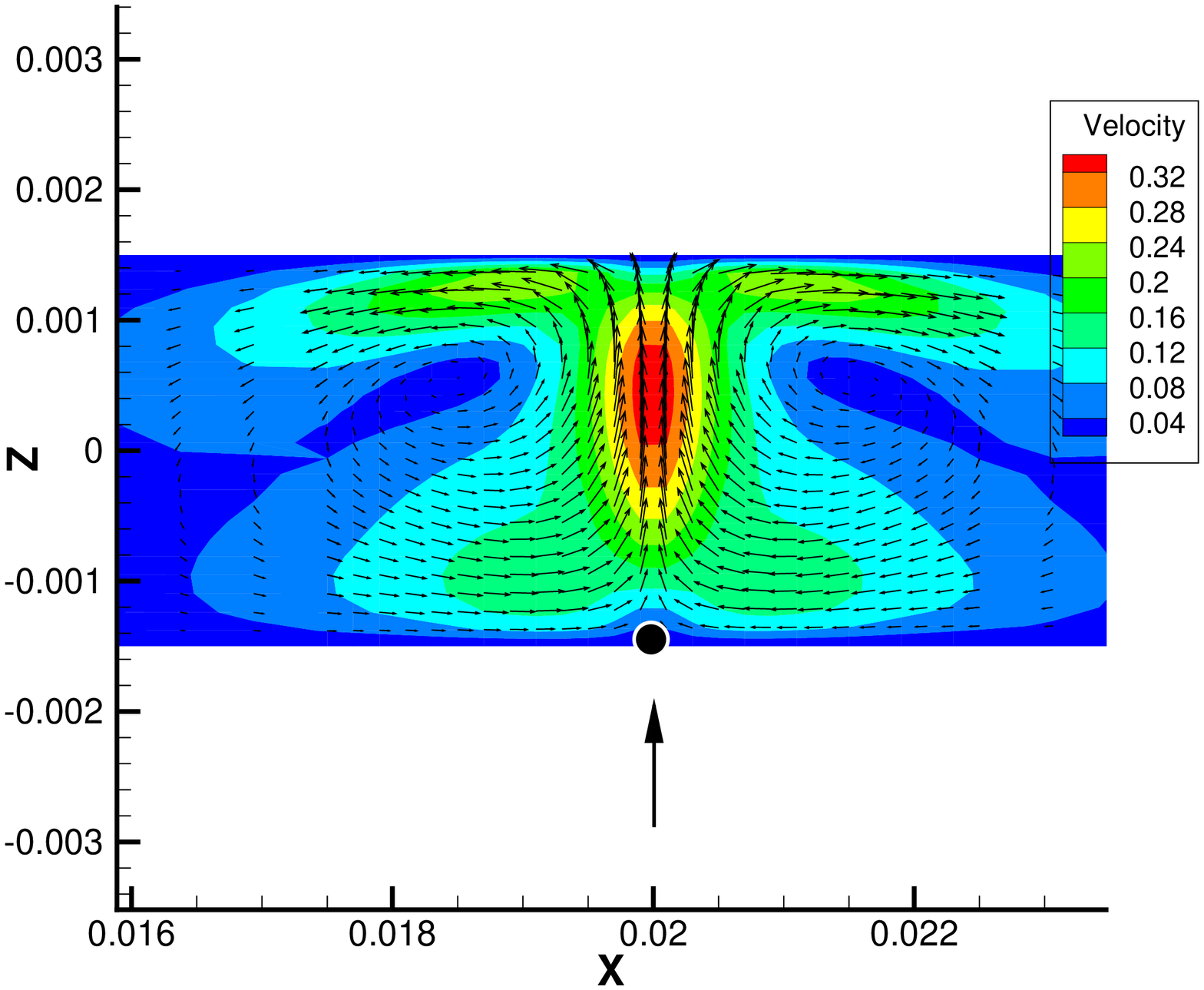}}
      \subfigure[]{\includegraphics[width=0.34\textwidth]{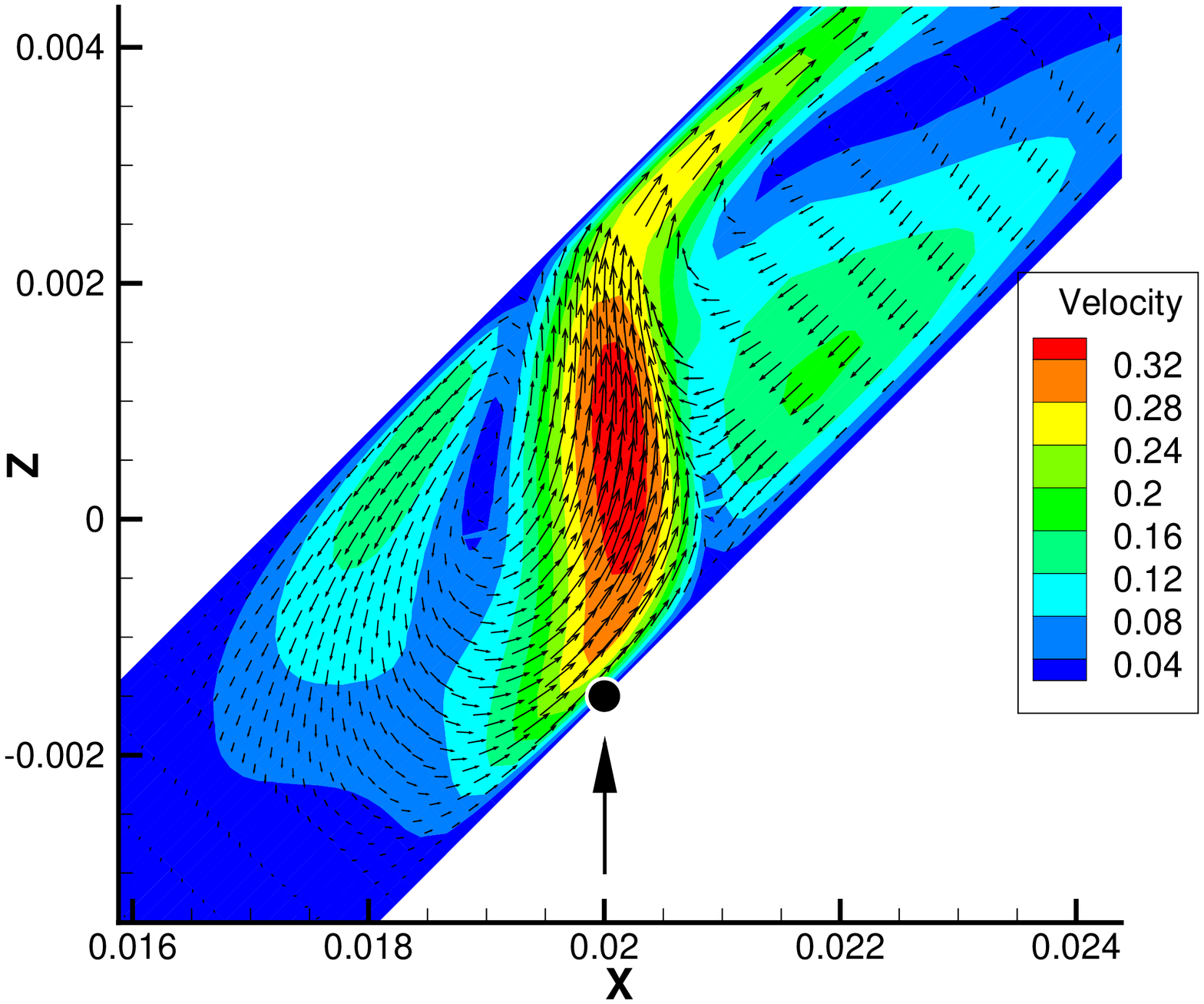}}
      \subfigure[]{\includegraphics[width=0.21\textwidth]{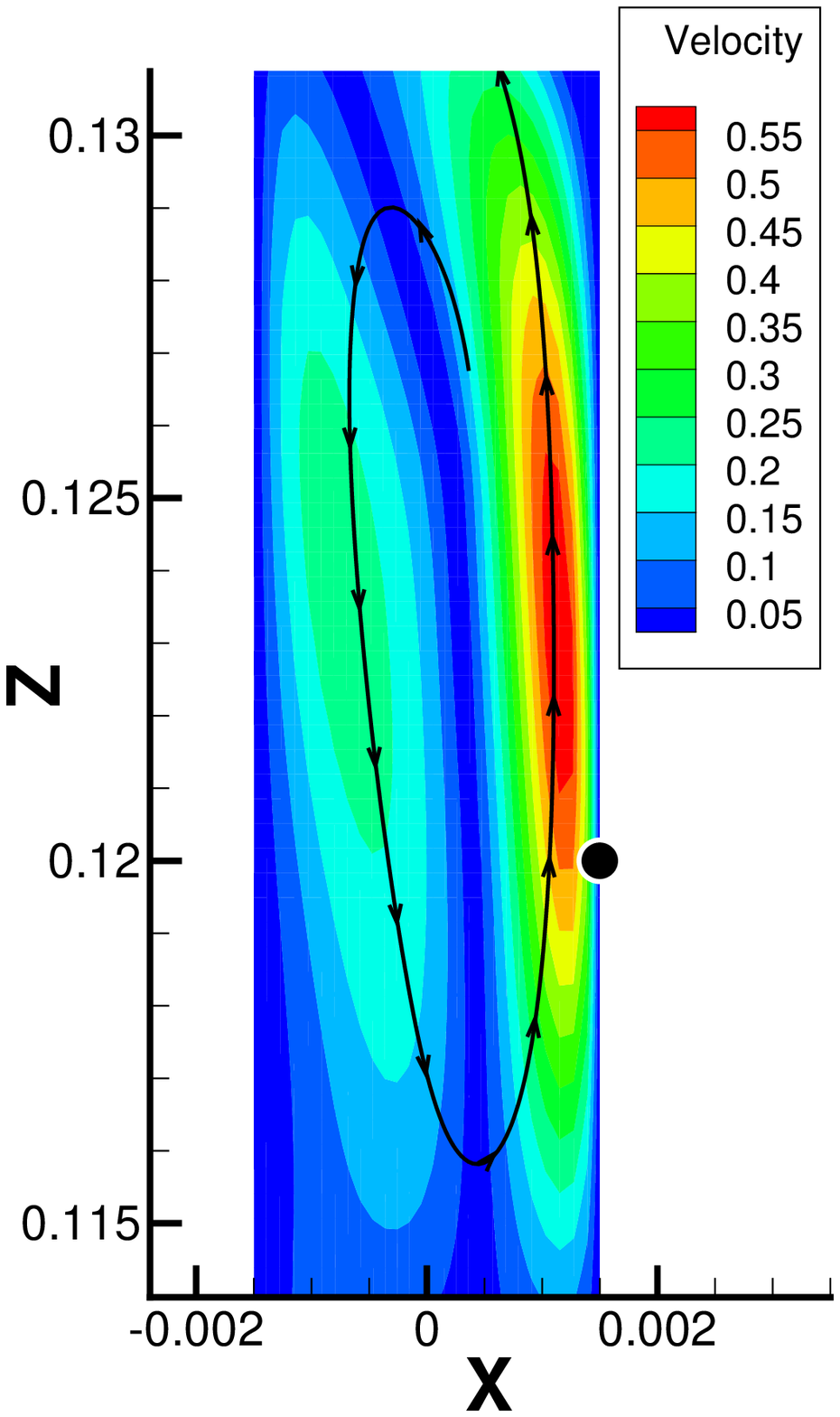}}}
\caption{The predicted acoustic streaming velocity vectors and magnitudes contours at the cutting plane in the blood vessel without an externally applied flow (blood flow=0) for different sonication angles $90^0$ (a), $45^0$ (b) and $0^0$ (c). Focal point ($\bullet$) is located on the blood vessel wall.}
\label{fig:AS_d3}
\end{figure}

\begin{figure}
\centering
\includegraphics[width=0.55\textwidth]{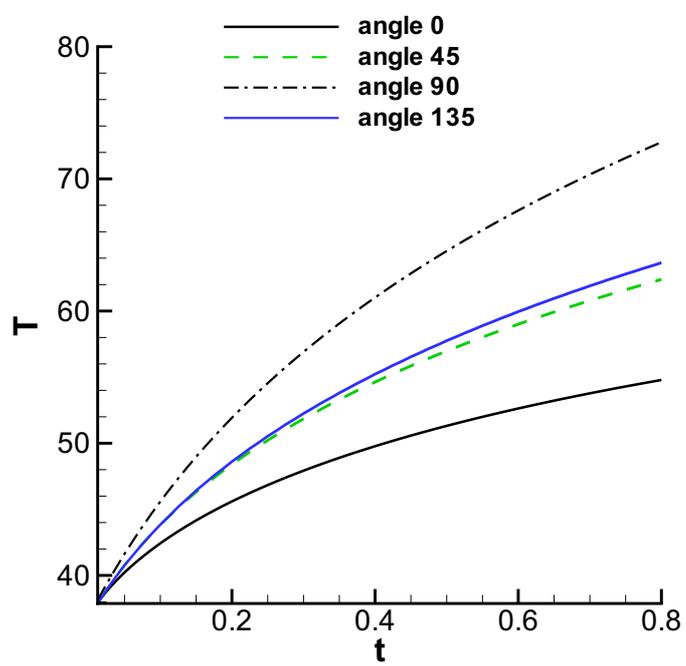}
\caption{The predicted temperature at the focal point as the function of time for different sonication angles.}
\label{fig:T_focus_time}
\end{figure}

\end{document}